\documentclass[sigconf]{acmart}
\AtBeginDocument{%
  \providecommand\BibTeX{{%
    \normalfont B\kern-0.5em{\scshape i\kern-0.25em b}\kern-0.8em\TeX}}}

\copyrightyear{2024}
\acmYear{2024}
\setcopyright{rightsretained}
\acmConference[ASSETS '24]{The 26th International ACM SIGACCESS Conference on Computers and Accessibility}{October 27--30, 2024}{St. John's, NL, Canada}
\acmBooktitle{The 26th International ACM SIGACCESS Conference on Computers and Accessibility (ASSETS '24), October 27--30, 2024, St. John's, NL, Canada}
\acmDOI{10.1145/3663548.3675637}
\acmISBN{979-8-4007-0677-6/24/10}
\usepackage{graphicx}
\usepackage{float}
\usepackage{caption}
\usepackage{subcaption}
\usepackage{longtable}
\usepackage{soul}
\usepackage{listings}

\begin{document}

\title[Vision-Based Assistive Technologies for People with Cerebral Visual Impairment]{Vision-Based Assistive Technologies for People with Cerebral Visual Impairment: A Review and Focus Study}

\author{Bhanuka Gamage}
\email{bhanuka.gamage@monash.edu}
\orcid{0000-0003-0502-5883}
\affiliation{%
  \institution{Monash University}
  \streetaddress{Wellington Rd}
  \city{Melbourne}
  \country{Australia}
}

\author{Leona Holloway}
\email{leona.holloway@monash.edu}
\orcid{0000-0001-9200-5164}
\affiliation{%
  \institution{Monash University}
  \streetaddress{Wellington Rd}
  \city{Melbourne}
  \country{Australia}
}

\author{Nicola McDowell}
\email{n.mcdowell@massey.ac.nz}
\orcid{0000-0001-6969-9604}
\affiliation{%
  \institution{Massey University}
  \city{Auckland}
  \country{New Zealand}
}

\author{Thanh-Toan Do}
\email{toan.do@monash.edu}
\orcid{0000-0002-6249-0848}
\affiliation{%
  \institution{Monash University}
  \streetaddress{Wellington Rd}
  \city{Melbourne}
  \country{Australia}
}

\author{Nicholas Price}
\email{nicholas.price@monash.edu}
\orcid{0000-0001-9404-7704}
\affiliation{%
  \institution{Monash University}
  \streetaddress{Wellington Rd}
  \city{Melbourne}
  \country{Australia}
}

\author{Arthur Lowery}
\email{arthur.lowery@monash.edu}
\orcid{0000-0001-7237-0121}
\affiliation{%
  \institution{Monash University}
  \streetaddress{Wellington Rd}
  \city{Melbourne}
  \country{Australia}
}

\author{Kim Marriott}
\orcid{0000-0002-9813-0377}
\email{kim.marriott@monash.edu}
\affiliation{%
  \institution{Monash University}
  \streetaddress{Wellington Rd}
  \city{Melbourne}
  \country{Australia}
}

\renewcommand{\shortauthors}{Bhanuka Gamage, et al.}

\begin{abstract}
Over the past decade, considerable research has investigated Vision-Based Assistive Technologies (VBAT) to support people with vision impairments to understand and interact with their immediate environment using machine learning, computer vision, image enhancement, and/or augmented/virtual reality.
However, this has almost totally overlooked a growing demographic: people with Cerebral Visual Impairment (CVI). 
Unlike ocular vision impairments, CVI arises from damage to the brain's visual processing centres.
Through a scoping review, this paper reveals a significant research gap in addressing the needs of this demographic. 
Three focus studies involving 7 participants with CVI explored the challenges, current strategies, and opportunities for VBAT.
We also discussed the assistive technology needs of people with CVI compared with ocular low vision. 
Our findings highlight the opportunity for the Human-Computer Interaction and Assistive Technologies research community to explore and address this underrepresented domain, thereby enhancing the quality of life for people with CVI.
\end{abstract}



\begin{teaserfigure}
  \includegraphics[width=0.95\textwidth,keepaspectratio]{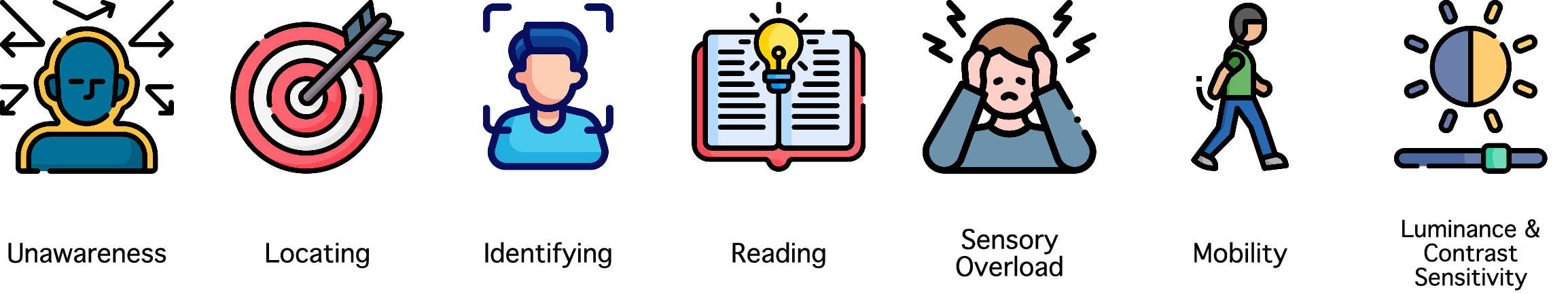}
    \centering
    \caption{Overview of challenges identified in the focus group discussions with people living with CVI. Icon source: \cite{flaticons}.}
    \Description{Seven icon-based images depicting the challenges: Unawareness, Locating, Identifying, Reading, Sensory Overload, Mobility, and Luminance & Contrast Sensitivity.}
    \label{fig:challenges}
\end{teaserfigure}


\maketitle

\section{Introduction}

\begin{figure*}[!t]
    \centering
    \includegraphics[width=14cm,keepaspectratio]{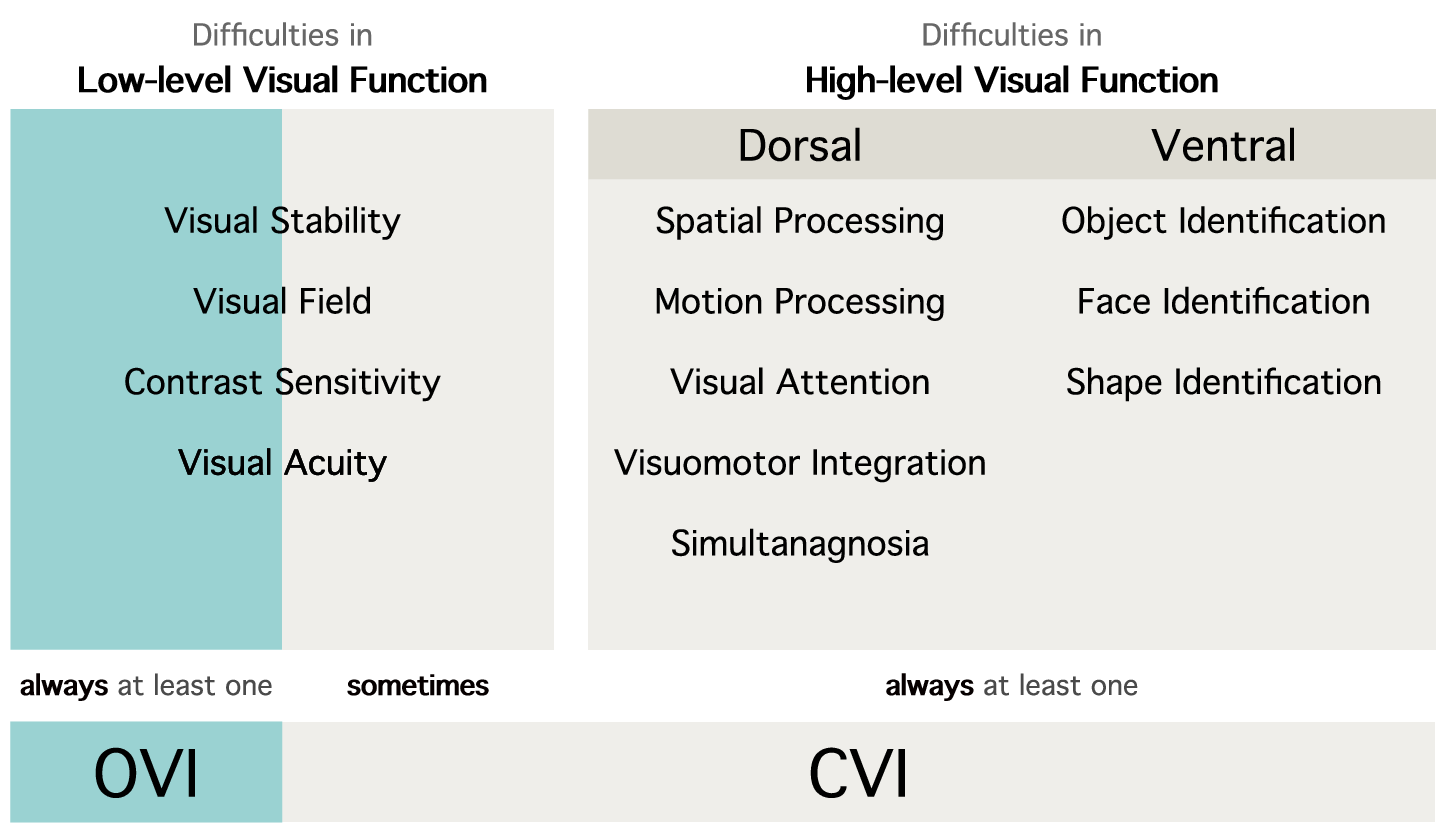}
    \caption{Overview of low-level and high-level visual difficulties for people with CVI. Note that this list is not exhaustive. Concepts depicted are extracted from \cite{lueck2015vision}.}
    \Description{Two squares labelled “Difficulties in Low-level Visual Function” and “Difficulties in High-level Visual Function”. Under the squares there is a horizontal bar indicating that at least one low-level visual difficulty is always present in people with OVI, and sometimes for people with CVI, while at least one high-level visual difficulty is always present for people with CVI. The low-level visual difficulties are as follows: Visual Stability, Visual Field, Contrast Sensitivity, Visual Acuity. The high level visual difficulties are subdivided into two sub-categories - Dorsal and Ventral. The dorsal difficulties are as follows: Spatial processing, Motion processing, Visual Attention, Visuomotor Integration, Simultanagnosia. The ventral difficulties are as follows: Object Identification, Face Identification, Shape Identification.}
    \label{fig:cvi-overview}
\end{figure*}

Vision is not solely a product of the eyes; it also relies on intricate processes of the brain.
\textbf{Cerebral Visual Impairment (CVI)} is a visual dysfunction that is distinct from Ocular Vision Impairment (OVI), and caused by injury or disruption to the brain's visual processing centres \cite{sakki2018there}. 
Like OVI, CVI can cause difficulties in low-level visual processing functions such as reducing visual acuity or the visual field. 
Unlike OVI, CVI also causes difficulties in high-level visual processing such as object or face recognition and control of visual attention (see Figure~\ref{fig:cvi-overview}).
CVI has emerged as the predominant cause of childhood vision impairment in developed nations \cite{sandfeld2007visual, hatton2007babies, matsuba2006long}, affecting around 30-40\% of visually impaired children \cite{americanpediatric}.
As these children transition into adulthood, CVI is poised to become the leading cause of vision impairment \cite{bosch2014low}.

Recent advancements in artificial intelligence, particularly in computer vision and multimodal large language models~\cite{vaswani2017attention, yin2023survey, wu2023multimodal, chatgpt} and in augmented and virtual reality \cite{ar_trend, ricci2023virtual, munoz2022augmented, wu2021towards, masnadi2020vriassist} have spawned a revolutionary new class of assistive technologies for people with vision impairment~\cite{li2022scoping, lee2021deep, assets2023}.
These include object recognition \cite{bashiri2018object, arora2019real, joshi2020yolo}, autonomous guidance systems~\cite{chen2020smart, otaegui2013argus},  visual guidance \cite{vision_guidance}, compensation for colour blindness \cite{vision_color_blind}, and visual noise cancellation through vision augmentation \cite{visual_noise}. 
We refer to these as \textbf{Vision-Based Assistive Technologies (VBAT)}, by which we mean personal wearable devices that incorporate AI-based computer vision technologies, including machine learning, computer vision, image enhancement, and/or augmented/virtual reality to enhance understanding and interaction with the immediate environment.

However, as we shall see (Section~\ref{section: review}) virtually all of this research has focused on the needs of people with OVI. 
This study addresses the primary question: \textbf{What is the current state of research at the intersection of CVI and VBAT, and what opportunities exist in this domain?}
% To explore this question, we begin by introducing CVI and then examine the potential opportunities of VBAT in this context.
Our main research contributions are:
\begin{itemize}
    \item \textbf{Scoping Review:} We conducted a scoping review \cite{munn2018systematic} into the current research concerning CVI and technologies for VBAT. This revealed a paucity of research and a predominant focus on understanding CVI rather than assistance. 
    \item \textbf{Opportunities for VBAT:} We conducted three focus groups with 7 participants with CVI (one of whom is an author). These identified 7 broad challenges, ranging from awareness of objects to sensory overload in face-to-face conversations, that people with CVI face. We then ideated the use of VBAT to address these challenges.
    \item \textbf{Comparison of the AT needs of people with CVI and those with ocular low vision (OLV):} Based on prior research and our focus groups, we have identified the similarities and differences between these two groups ranging from single modality preferences to the impacts of visual complexity.
\end{itemize}
More fundamentally, as one of the first research papers to consider assistive technology for people with CVI, we hope that our paper
\begin{itemize}
  \item \textbf{Raises Awareness of CVI in HCI and Assistive Technology (AT) Communities} by highlighting their needs and clarifying that these are not the same as those of people with OLV and that researchers should consider and report them as a distinct cohort.
\end{itemize}
\section{Background - Cerebral Visual Impairment (CVI)}
\label{sec: cvi}
Cerebral Visual Impairment (CVI) is the result of damage to the brain’s visual processing centres, rather than physical damage to the eye \cite{lueck2015vision, roman2007cortical}. 
CVI is primarily observed in individuals with neurological conditions like cerebral palsy, stroke, or traumatic brain injury.
People with CVI face issues with interpreting and processing visual information, including difficulties in visual recognition, perception, understanding of their visual environment, and maintaining visual attention. 

The terminology and definitions related to this condition, including Cerebral Visual Impairment \cite{philip2014identifying, lueck2015vision}, Cortical Vision Impairment \cite{whiting1985permanent, good1994cortical, roman2007cortical}, and Neurological Vision Impairment \cite{trobe2001neurology}, vary subtly and are a subject of ongoing debate %among researchers, medical professionals, and practitioners 
~\cite{martin2016cerebral, mcdowell2020cvi}. 
We will use Cerebral Visual Impairment (CVI) to encompass a range of visual processing challenges resulting from damage to the brain's visual processing centres.

Ocular Visual Impairment (OVI) refers to vision impairments arising from problems within the eye or its associated structures, such as the retina, optic nerve, or cornea \cite{martin2016cerebral}. 
OVI examples include conditions like cataracts, glaucoma, and macular degeneration, which can cause a range of vision impairments from partial (low-vision) to total blindness.

\subsection{CVI Diagnosis and Prevalence}
\label{sec:morecvichildren}

In the last decade, our understanding of CVI has significantly expanded \cite{lueck2015vision, roman2007cortical}. 
Consequently, there has been an upsurge in the diagnosis of CVI among children; this can be attributed to the dissemination of knowledge regarding its primary causes, especially premature birth, among other factors \cite{bennett2020neuroplasticity, dutton2004association, good2001recent, kozeis2010brain, taylor2009differential}. 
CVI is frequently under-diagnosed, primarily because of coexisting brain damage causing other physical and cognitive impairments. 
In some instances, CVI might be mistaken for conditions like autism, learning disabilities, or behavioural challenges, further complicating accurate diagnosis \cite{swift2008cortical, williams2021cerebral}.

Numerous studies have focused on identifying children with CVI. 
Williams et al. \cite{williams2021cerebral} screened a substantial cohort of children and found a CVI prevalence rate of 3.4\% among those in mainstream classrooms. 
In addition to the large prevalence of CVI in children, many adults may grapple with undiagnosed CVI. %. 
Furthermore, adults can also develop CVI later in life, often stemming from factors such as strokes, traumatic brain injuries, %degenerative brain diseases, multiple sclerosis, Parkinson's disease, 
and other neurological conditions \cite{lehmann2011basic, zhang2006homonymous}. 

At present, assessments and classifications of visual capability focus on OVI and primarily rely on criteria related to visual acuity and visual field. 
People with CVI may have normal visual acuity, but struggle with higher-level visual functions~\cite{hyvarinen1995considerations, saidkasimova2007cognitive}.
For instance, a child with CVI may have difficulty recognising a parent only when in a crowded room, or finding a particular toy only when it's mixed with other objects \cite{dutton2003cognitive}. 
This presents clinicians with challenges when diagnosing people with CVI, because the existing framework does not sufficiently address their needs.
It also means that many individuals with CVI do not meet the legal definition of blindness as per current standards \cite{kran2019cerebral}. 
This underscores the necessity for a broader understanding of CVI along with advocacy for a revision of the definitions of vision impairment to encompass people with CVI. 

\subsection{Low-Level Functional Vision vs. High-Level Functional Vision}
\label{sec:low-high-function}

People with CVI commonly experience difficulties in both low-level and high-level functional vision \cite{bennett2019assessment, chandna2021higher}.
Low-level functional vision encompasses fundamental aspects of visual perception like visual acuity, contrast sensitivity, visual field, and visual stability. These are the type of perceptual issues faced by people with OVI.

Conversely, high-level functional vision pertains to how individuals interpret and respond to visual information, including processes like visual recognition, comprehension, visual attention, and interaction with the environment. 
Studies indicate that these high-level vision impairments are associated with damage in two critical visual processing pathways in the brain--the Dorsal Stream and Ventral Stream \cite{bennett2020neuroplasticity}.

The two-stream model proposed by Goodale \cite{goodale2013separate}, consisting of the `dorsal stream'--connecting the occipital to the parietal cortex and the `ventral stream'--connecting the occipital to the inferior temporal cortex, offers a valuable framework for understanding higher-order visual processing difficulties in CVI \cite{haxby1991dissociation, mishkin1983object}.
Dysfunction in the dorsal stream typically manifests as vision impairments related to spatial and motion processing, difficulties in visual attention, challenges in visuomotor integration, and Simultanagnosia\footnote{Simultanagnosia is the inability to perceive more than one object at a time causing difficulty perceiving the entire visual scene.}\cite{bennett2020neuroplasticity, frcophth2009dorsal}. 
Meanwhile, damage along the ventral visual processing stream is associated with difficulties in object identification~\cite{bennett2020neuroplasticity, goodale2013separate, haxby1991dissociation}, such as recognising faces and shapes~\cite{andersson2006vision, houliston1999evidence}. 
However, dorsal and ventral stream difficulties often co-occur, suggesting that damage cannot be localised to a single brain area \cite{bennett2018assessing, dutton2011structured, macintyre2010dorsal}.
Figure \ref{fig:cvi-overview} provides an overview of the types of visual difficulties for people with CVI.

Recent advancements have provided tools to identify high-level visual function difficulties in people with CVI \cite{chandna2021higher, mcdowell2022using}.
These include questionnaires for initial assessment \cite{chandna2021higher}, and an iPad app that induces visual crowding \cite{mcdowell2020cvi, mcdowell2023validation}.

\section{Related Work}
\label{sec: relatedworks}

In this section, we provide a brief introduction to Vision-Based Assistive Technologies (VBAT) and review relevant prior research into CVI. 

\subsection{Vision Based Assistive Technologies (VBAT)}
In this paper we focus on the use of Vision-Based Assistive Technologies (VBAT). These are devices incorporating machine learning, computer vision, image enhancement, and/or augmented/virtual reality, to enhance understanding and interaction with the immediate environment. A large body of research within the fields of HCI and AT has been dedicated to aiding people with OVI understand their environment \cite{patel2020assistive, simoes2020review, kuriakose2020multimodal, khan2021insight, valipoor2022recent, bashiri2018object, arora2019real, joshi2020yolo, chen2020smart, otaegui2013argus}.
These studies can be categorised into two primary areas: studies focusing on audio or tactile solutions for people who are blind (Ocular Blindness) \cite{abdolrahmani2021towards, kaul2021mobile, bhatia2022audio, villamarin2021haptic, yasmin2020haptic, srija2020raspberry}, and studies focusing on visual devices for people with low vision (Ocular Low Vision - OLV) \cite{zhao2015foresee, lang2020augmented, zhao2016cuesee, zhao2019designing, zhao2020effectiveness}.

In recent years, significant advancements have been made in artificial intelligence, particularly in fields such as computer vision and multimodal large language models~\cite{vaswani2017attention, yin2023survey, wu2023multimodal, caffagni2024r, chatgpt}.
Consequently, there has been a growing interest in developing AI-based assistive technologies for people with OVI \cite{assets2023, li2022scoping, lee2021deep}. 
These technologies span from basic object recognition \cite{bashiri2018object, arora2019real, joshi2020yolo}, to advanced autonomous guidance systems~\cite{chen2020smart, otaegui2013argus}, integrated into various devices such as smart glasses~\cite{guarese2021cooking, islam2020design, lin2020smart} and robotic dogs~\cite{hong2022development, bruno2019development}.
Additionally, numerous commercially available AI smart devices have emerged, offering a variety of assistive functionalities ~\cite{seeingai, orcam, envisionglasses, straptech, nueyes, visionbuddy}.
Several studies have reviewed these devices and technologies, providing insights into their capabilities and applications \cite{assets2023, patel2020assistive, simoes2020review, kuriakose2020multimodal, khan2021insight, valipoor2022recent}.
In particular, Gamage et al. \cite{assets2023} reviewed smart devices employing AI-based computer vision to assist people with OVI in comprehending their surroundings and identified 646 studies published between 2020 and 2022.

Additionally, with recent technological advancements in augmented and virtual reality (AR/VR) \cite{ar_trend}, research has explored the applications and effectiveness of AR/VR for people with OVI \cite{ricci2023virtual, munoz2022augmented, wu2021towards, masnadi2020vriassist, pur2023use}. This technology has diverse applications, including visual guidance \cite{vision_guidance}, compensation for colour blindness \cite{vision_color_blind}, and visual noise cancellation through vision augmentation \cite{visual_noise}. 
In most cases AR is combined with AI-based technologies.

Pur et al. \cite{pur2023use} conducted a review of 16 studies investigating AR/VR for visual field expansion and visual acuity improvement in people with OLV. 
They found that AR/VR devices can enhance the visual field and acuity, with the majority of studies utilising AR technology. 
Commercial head-mounted displays (HMDs) were commonly used across the studies.
Building upon the visual capabilities of HMDs in AR/VR, Li et al. \cite{li2022scoping} conducted a scoping review exploring the use of HMDs as assistive and therapeutic devices for people with visual impairments.
Their analysis revealed a growing body of research utilising HMDs for visual assistance and therapy.
AR was predominantly used for visual assistance, while VR was employed for therapeutic purposes. 

\subsection{Assistive Technologies and CVI}

Many of the reviews conducted in the field of CVI predominantly focus on the medicine and education domains \cite{philip2014identifying, delay2023interventions, mcdowell2023review, mcconnell2021assessments}.
Philip and Dutton \cite{philip2014identifying} provided an overview of the features of CVI and practical management strategies for supporting children with this condition.
They highlighted the diverse causes of CVI and its common occurrence in children with cerebral palsy. 
Additionally, they emphasised the importance of employing a structured approach to history-taking for easy identification of CVI.

Delay et al. \cite{delay2023interventions} conducted a scoping review to investigate intervention studies for children with CVI and found limited evidence in this area. 
They observed that most intervention studies have low-level evidence, underscoring the need for high-quality, controlled intervention studies to inform evidence-based practice for families and clinicians.

McDowell \cite{mcdowell2023review} conducted a literature review to inform the development of a practice framework for supporting children with CVI. 
The review also highlighted the current reliance on approaches and strategies developed for children with ocular visual impairments to support children with CVI.

McConnell et al. \cite{mcconnell2021assessments} conducted a systematic review of assessments used to investigate and diagnose CVI in children. 
They found a lack of standardised approaches among clinicians, with diagnosis often relying on a `diagnosis of exclusion'\footnote{`Diagnosis of exclusion' refers to systematically ruling out other possible conditions based on observed symptoms.} method.

Research on the assistive technology needs of people with CVI is sparse \cite{broadening_our_view}, with the majority of existing studies focusing on specific case studies involving children \cite{lane2023case, furze2018integrating} or reports from parents \cite{goodenough2021bridging, lupon2018quality}.
Considering the significance of the visual information for people with CVI, the question arises as to whether VBAT technologies can be tailored to address their specific needs and work as assistive technologies.
Could these devices leverage a combination of machine learning, image enhancement, and augmented reality to provide real-time assistance? 
For example, VBAT could be used to fade out unwanted details when locating a pair of scissors, finding friends in a crowded venue, or finding a shop on a busy street.
To date, there have been no reviews conducted at the intersection of VBAT and CVI, resulting in a notable gap in understanding the potential benefits of such technologies for people with CVI. 
Moreover, there is a lack of insight into the main challenges and assistive technology needs of people with CVI that these devices could potentially address. 
These are the primary issues we aim to investigate and address in this paper.

\begin{table*}[h]
  \centering
  \caption{Scoping review keywords and criteria}
  \begin{subtable}{0.42\textwidth}
    \centering
            \caption{Keywords}
            \begin{tabular}{|p{4.15cm}|p{2.8cm}|}
            \hline
                \textbf{Cerebral Visual Impairment} & \textbf{Technologies for \newline VBAT} \\ \hline
                "cerebral visual impairment" & "computer vision" \\ 
                "cortical visual impairment" & "artificial intelligence" \\ 
                "neurological vision impairment" & "machine learning" \\ 
                ~ & "image enhancement" \\ 
                ~ & "augmented reality" \\ 
                ~ & "virtual reality" \\ 
                ~ & "mixed reality" \\ \hline
            \end{tabular}
            \label{tab: keywords}
  \end{subtable}
  \hspace{3mm}
  \begin{subtable}{0.5\textwidth}
    \centering
        \caption{Inclusion and Exclusion Criteria}
        \begin{tabular}{|p{9cm}|}
        \hline
        \textbf{Inclusion}                                                                  \\ \hline
        Papers focusing on CVI and technologies in Table \ref{tab: keywords}, irrespective of whether CVI participants are involved. \\
        Papers with CVI participants and technologies in Table \ref{tab: keywords}, regardless if focus is on CVI.   \\
        \hline
        \textbf{Exclusion}                                                                  \\ \hline
        Review papers, Theses or Non-English papers. \\                
        Papers with CVI only mentioned in the reference list. \\
        Papers without any technologies in Table \ref{tab: keywords}. \\
        \hline
        \end{tabular}
        \label{tab:criteria}
  \end{subtable}
  \label{main-table}
\end{table*}

\section{Scoping Review}
\label{section: review}

The first phase in our research was to conduct a scoping review, examining the intersection of CVI with technologies for VBAT. 
A scoping review was chosen because it is well-suited for examining research practices in a specific topic, identifying knowledge gaps, and categorising available evidence within a given field~\cite{munn2018systematic}.

Our search was structured around two search criteria: Cerebral Visual Impairment and Technologies for VBAT.
The keywords, displayed in Table~\ref{tab: keywords}, were combined using "OR" within the same column and "AND" across columns during the search process. 
We started by conducting a full-text search across four databases: Google Scholar, Scopus, Web of Science, and PubMed. 
The search captured papers published up to January 2, 2024. It yielded 595 papers from Google Scholar, 210 from Scopus, 9 from Web Of Science, and 5 from PubMed.
After consolidating and removing duplicates, a title and abstract screening process was conducted by one researcher with consultation from other members of the research team, leading to 67 papers for full-text review.
Two researchers then reviewed the full texts using the criteria in Table~\ref{tab:criteria}, discussing and resolving any conflicts to identify a set of 14 papers at the intersection of CVI and technologies for VBAT.
The search was subsequently refreshed on March 25, 2024, finding three more papers.

\subsubsection*{\textbf{Data Extraction:}}

The following information was recorded for each paper:
\begin{itemize}
    \item \textbf{Type of Study:} The primary focus of the study: Diagnosis (aiming to diagnose CVI), Understanding CVI (focused on understanding CVI), Simulation (simulating CVI conditions), Assistance (designed to help users during the actual use of the system, such as providing support for daily activities), or Rehabilitation (focused on providing therapeutic interventions or exercises to aid in the recovery and improvement of visual functions). These categories were determined by the two researchers conducting the full text review.
    \item \textbf{Type of Technology:} Specific type of technologies employed in the study.
    \item \textbf{Age Group:} The age groups of the participants in the study; adults (over 18 years), children (below 18 years), both or unclear.
    \item \textbf{Involvement of CVI Participants:} Whether people with CVI were part of the study and, if so, whether their participation was during the design phase, evaluation phase, or both.
    \item \textbf{Year and Other Details:} Publication year and additional information, including publication venue, and authors' names.
    
\end{itemize}

\subsection{\textbf{Results}}

This section offers a summary of the main findings; Table \ref{tab:full-paper-list} (in the Appendix) gives the full dataset.

\subsubsection*{Type of Study:}
\begin{figure*}[hbt!]
    \centering
    \begin{subfigure}{0.48\textwidth}
        \centering
        \includegraphics[width=\textwidth]{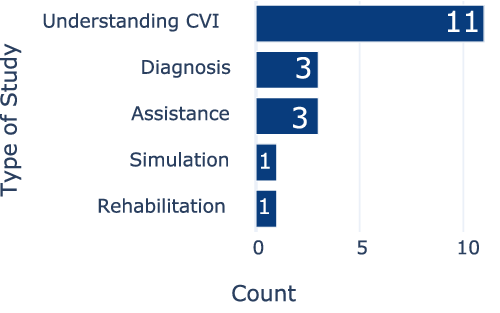}
        \caption{Type of Study}
        \Description{Horizontal bar chart showing the count of papers for each type of study. The types are sorted from highest to lowest count and are as follows: Understanding CVI 11, Diagnosis 3, Assistance 3, Simulation 1, Rehabilitation 1.}
        \label{fig:typeofstudy}
    \end{subfigure}
    \hfill
    \begin{subfigure}{0.48\textwidth}
        \centering
        \includegraphics[width=\textwidth]{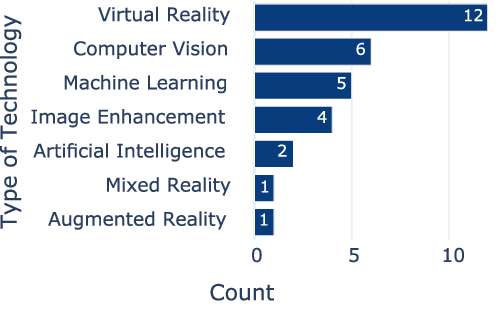}
        \caption{Type of Technology}
        \Description{Horizontal bar chart showing the count of papers for each type of technology. These are sorted from highest to lowest count and are as follows: Virtual Reality 12, Computer Vision 6, Machine Learning 5, Image Enhancement 4, Artificial Intelligence 2, Mixed Reality 1, Augmented Reality 1.}
        \label{fig:technologies}
    \end{subfigure}
    \vskip\baselineskip
    \begin{subfigure}{0.32\textwidth}
        \centering
        \includegraphics[width=\textwidth]{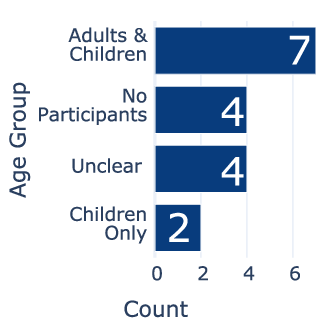}
        \caption{Age Group}
        \Description{Horizontal bar chart showing the count of papers for participant age groups. The counts are as follows: Adults \& Children 7, No participants 4, Unclear 4, Children only 2.}
        \label{fig:age-group}
    \end{subfigure}
    \begin{subfigure}{0.32\textwidth}
        \centering
        \includegraphics[width=\textwidth]{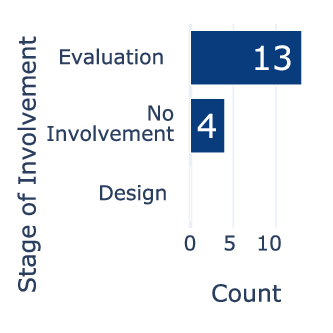}
        \caption{Stage of Involvement}
        \Description{Horizontal bar chart showing the count of papers for stage of involvement. The counts are as follows: Evaluation 13, No Involvement 4, Design 0.}
        \label{fig:involvement-stage}
    \end{subfigure}
    \begin{subfigure}{0.32\textwidth}
        \centering
        \includegraphics[width=\textwidth]{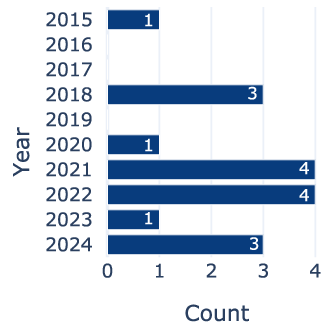}
        \caption{Publication Year}
        \Description{Horizontal bar chart showing the count of papers for publication year from 2015 to 2024. There is an increase over time. The counts are as follows: 2015 1, 2016 0, 2017 0, 2018 3, 2019 0, 2020 1 , 2021 4, 2022 4, 2023 1, 2024 3.}
        \label{fig: cvi-papers-year}
    \end{subfigure}
    \hfill
    \caption{Summary of findings from the Scoping Review}
    \Description{Five plots depicting summaries of scoping review findings titled 3A Publication Year, 3B Age Group, 3C Stage of Involvement, 3D Type of Study, and 3E Type of Technology.}
    \label{fig:review-2-summary}
\end{figure*}

Figure \ref{fig:typeofstudy} displays the distribution of study types extracted from the 17 papers. 
Some papers fell into multiple categories, hence the total count exceeds 17. 

% Assistive Technologies
We found only three papers that were directly relevant to the use of VBAT technologies to assist people with CVI. 
Birnbaum et al. \cite{birnbaum2015enhancing} suggested that presenting high contrast and low spatial frequency visual stimuli could increase visual awareness for people with CVI. 
They also suggested that augmented/virtual reality headsets could modify real-time visual input to enhance visual detection; 
however, this was not implemented.

The other two assistance studies were not specifically focused on addressing concerns related to people with CVI \cite{pitt2023strategies, lorenzini2021personalized}.
Pitt and McCarthy \cite{pitt2023strategies} identified strategies for highlighting items within visual scene displays to support augmentative and alternative communication access.
They identified four methods for highlighting items: contrast based on light and dark, contrast based on colour, outline highlighting, and the utilisation of scale and motion.
In their future directions, they suggested the potential application of these strategies for people with CVI.
Lorenzini et al. \cite{lorenzini2021personalized} conducted a study with the objective to gather insights to reduce the likelihood of device abandonment when using portable head-mounted displays for telerehabilitation.
While the primary focus of their study was OLV, participants with CVI were included in the study. 
Despite this inclusion, the study does not provide specific findings or discussions related to CVI.

% Diagnosis
Of the other studies, three focused on diagnosing CVI: Soni and Waoo \cite{soni2023convolutional} demonstrated the effectiveness of a convolutional neural network model for identifying and gaining insights into CVI.
Bennett and colleagues \cite{manley2022assessing, bambery2022virtual} developed two novel virtual reality based visual search tasks to objectively assess higher order processing abilities in CVI. 
% Understanding CVI
Using the same tasks, they conducted many studies to expand the understanding of CVI.
Ten studies identified for understanding CVI in the review were from Bennett and colleagues \cite{bennett2018virtual, bennett2018assessing, bennett2020alpha, pamir2021visual, bennett2021visual, federici2022altered, bambery2022virtual, zhang2022assessing, tambala2024abstract, da2024abstract} while the eleventh study centered on using machine learning models to understand gaze patterns of people with CVI \cite{avramidis2024evaluating}.

\subsubsection*{Type of Technology:}

Figure \ref{fig:technologies} offers an overview of the technologies employed in the examined papers. 
Virtual reality emerged as the predominant technology, largely due to the contributions by Bennett and colleagues. 
However, focusing on the subset of the three papers utilising VBAT papers for assistance, computer vision, image enhancement, and machine learning were consistently proposed or employed. 
However, with only three papers, it is evident that these technologies are underutilised, presenting an opportunity for future studies to address this gap.

\subsubsection*{Age Group}

In the 7 studies including both adults and children (see Figure \ref{fig:age-group}), ages ranged from 7 to 28 years old. 
The focus on the paediatric and young adult population is understandable given the higher CVI prevalence in children (refer to Section \ref{sec:morecvichildren}). 
However, future studies should include participants across all age groups to better understand the diverse strategies employed by people with decades of experience with CVI.

\subsubsection*{Involvement of CVI Participants:}

As depicted in Figure \ref{fig:age-group}, 13/17 papers incorporated CVI participants in the studies.
However, participant involvement was primarily focused on understanding CVI and diagnosis.
Such involvement in medical research is common practice, and in the context of these papers, participants were predominantly engaged for the purpose of validating hypotheses or effectiveness of diagnosis. 
None of the studies involved participants in the design or requirement gathering stages (as shown in Figure \ref{fig:involvement-stage}).

\subsubsection*{Year and Other Details:}

Figure \ref{fig: cvi-papers-year} shows an increasing publication rate with time, aligning with observations made by other CVI researchers \cite{lueck2015vision, roman2007cortical}. 
\section{Focus Group Discussion}

The second phase of our research involved conducting focus group discussions with people living with CVI. 
The main aim was to understand the difficulties and challenges, explore their current strategies, and identify potential opportunities for VBAT devices to assist them.

\subsection{Participants}

Three focus group discussions were held over Zoom with 7 participants with CVI.
One of the participants was also a co-author of the paper. 
Each group included at least 2 CVI participants and lasted approximately two hours. 
Participants received a \$100 USD gift card as compensation for their time. 
All 7 participants had been diagnosed with CVI, with varying characteristics. 
Table \ref{tab:participant_details} provides a summary of the demographic details of the participants, including age, gender, age of onset, and age of diagnosis.

\subsection{Procedure}

The focus group discussion had 3 parts:

\subsubsection*{\textbf{Understanding the challenges faced by people living with CVI\@:}}
The focus groups commenced with each participant sharing details about their specific CVI condition.
The discussion then explored the difficulties and challenges they encounter in day to day life.
Participants also described the current strategies utilised to mitigate these challenges.

\begin{table}[tbp]
    \centering
    \caption{Details of the participants with CVI: Participant (P - Participant, RP - Researcher \& Participant), Age Group, Age of Onset, Age of Diagnosis, Other Neurological Conditions (CP - Cerebral Palsy, N - None)}
    \label{tab:participant_details}
    \begin{tabular}{|p{0.3cm}|p{1cm}|p{1cm}|p{0.9cm}|p{1.3cm}|p{1.7cm}|}
    \hline
        \# & \textbf{Age Group} & \textbf{Gender} & \textbf{Age of \newline Onset} & \textbf{Age of \newline Diagnosis} & \textbf{Other \newline Neurological \newline Conditions}\\ \hline
        P1 & 45 to 54 & Female & Birth & 45 & N \\ \hline
        P2 & 55 to 64 & Female & 58 & 59 & N\\ \hline
        P3 & 55 to 64 & Female & Birth & 56 & N\\ \hline
        P4 & 25 to 34 & Female & 23 & 30 & CP\\ \hline
        P5 & 55 to 64 & Male &  57 & 57 & N\\ \hline
        P6 & 18 to 24 & Male &  1 & 1 & CP\\ \hline
        RP & 35 to 44 & Female & 16 & 32 & N\\ \hline
    \end{tabular}
\end{table}
\subsubsection*{\textbf{Potential opportunities and solutions for VBAT devices:}}
\begin{figure*}[h]
    \centering
    \begin{subfigure}{0.24\textwidth}
        \centering
        \includegraphics[width=\textwidth]{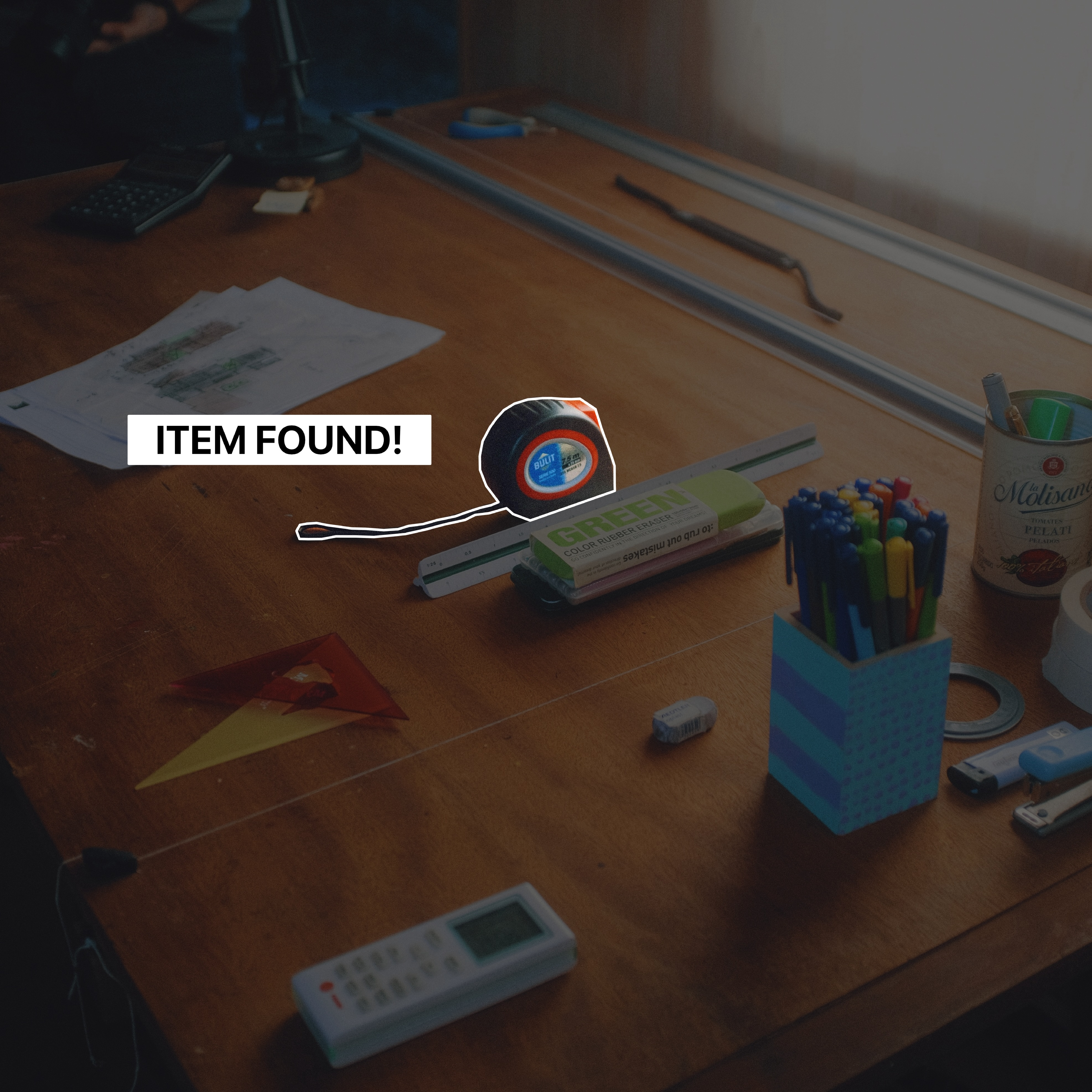}
        \caption{Utilising highlighting to locate a measuring tape. Original image: \cite{unsplash_measure_tape}}
        \Description{Photograph of a measuring tape on a table surrounded by rulers, pens, pen-holders, and a remote. The measuring tape is highlighted with a white border, and the rest of the image has a black tint. Beside the measuring tape is a white rectangle with black text saying "Item Found!".}
        \label{fig:image1}
    \end{subfigure}
    \hfill
    \begin{subfigure}{0.24\textwidth}
        \centering
        \includegraphics[width=\textwidth]{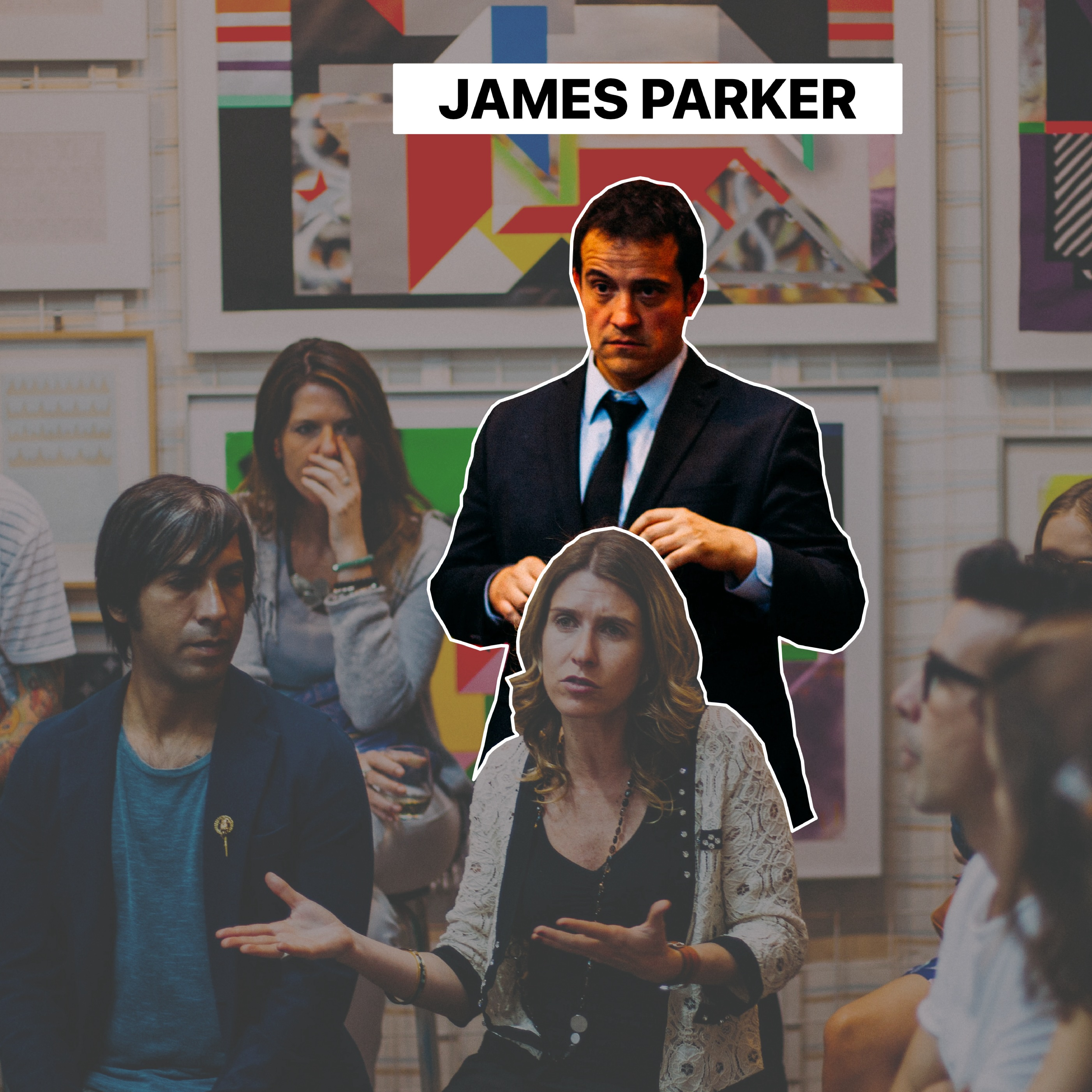}
        \caption{Utilising highlighting to identify an individual in a group. Original image: \cite{unsplash_people_room}}
        \Description{Photograph of seven people. A person in the second row is highlighted with a white border, and the rest of the image has a black tint. Above the head of the highlighted person is a white rectangle with black text saying "James Parker".}
        \label{fig:image2}
    \end{subfigure}
    \hfill
    \begin{subfigure}{0.24\textwidth}
        \centering
        \includegraphics[width=\textwidth]{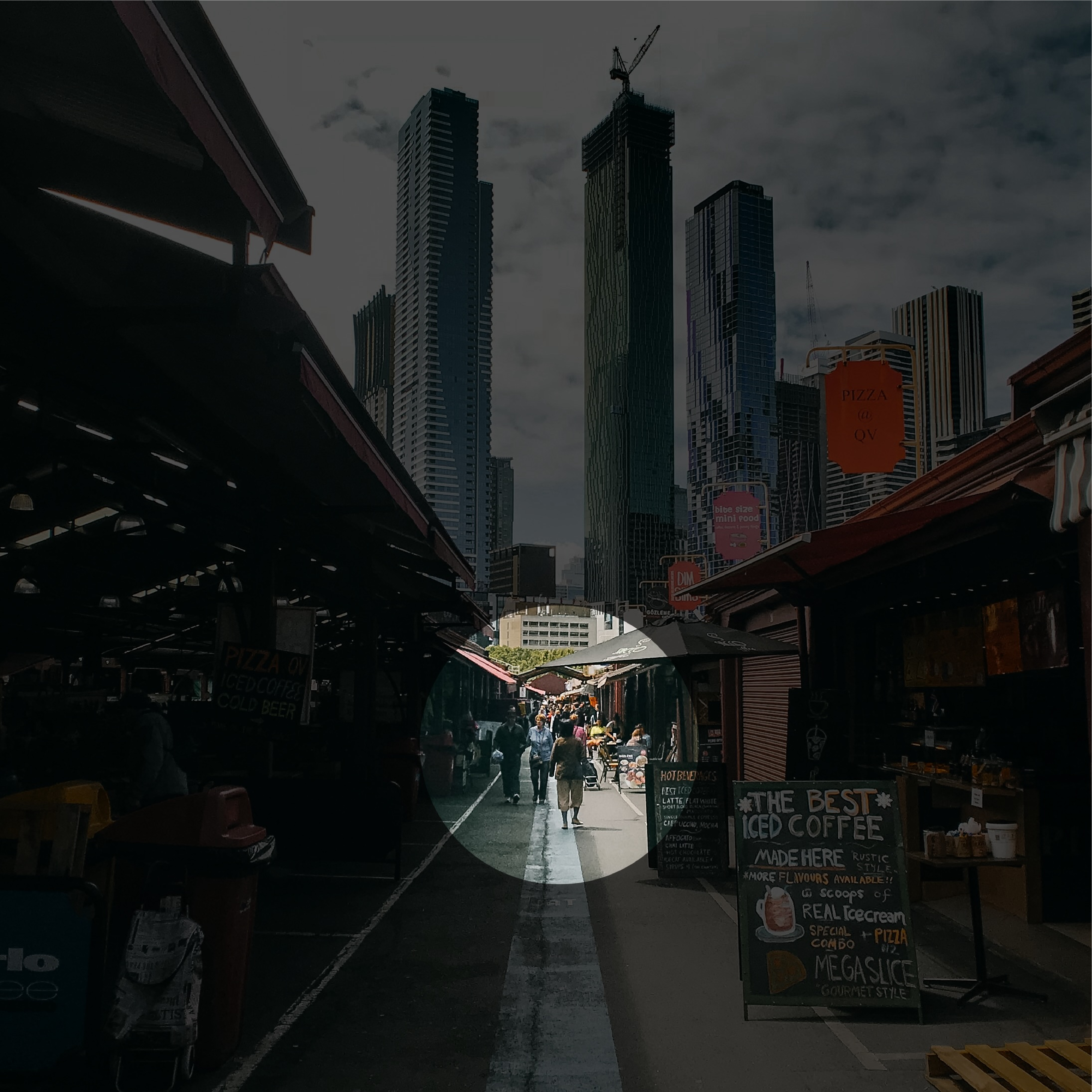}
        \caption{Incorporating tunnel vision to focus attention to the centre. Original image: \cite{unsplash_melb}}
        \Description{Photograph of an alleyway at a city market with tall skyscrapers in the background. People are walking and there are shops along the alleyway. The image has a black faded tint, except for a circular area at the centre.}
        \label{fig:image3}
    \end{subfigure}
    \hfill
    \begin{subfigure}{0.24\textwidth}
        \centering
        \includegraphics[width=\textwidth]{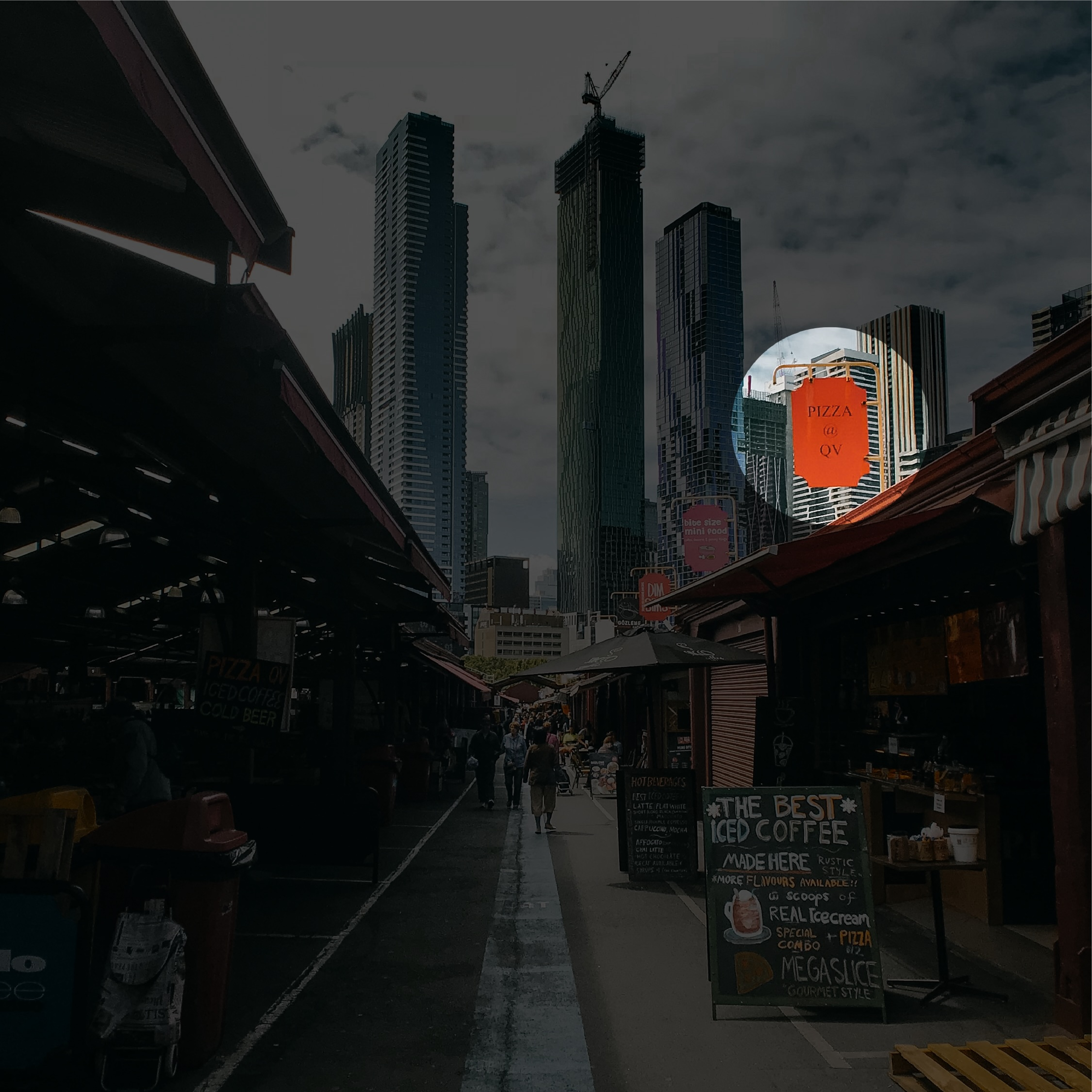}
        \caption{Implementing tunnel vision with bright colours to enhance focus. Original image: \cite{unsplash_melb}}
        \Description{Same photograph as Figure 4C of an alleyway at a city market. The image has a black faded tint, except for a circular area towards the top right, directing the user's attention to a bright orange shop name board.}
        \label{fig:image4}
    \end{subfigure}
    
    \vskip\baselineskip
    
    \begin{subfigure}{0.49\textwidth}
        \centering
        \begin{subfigure}{0.48\linewidth}
            \centering
            \includegraphics[width=\linewidth]{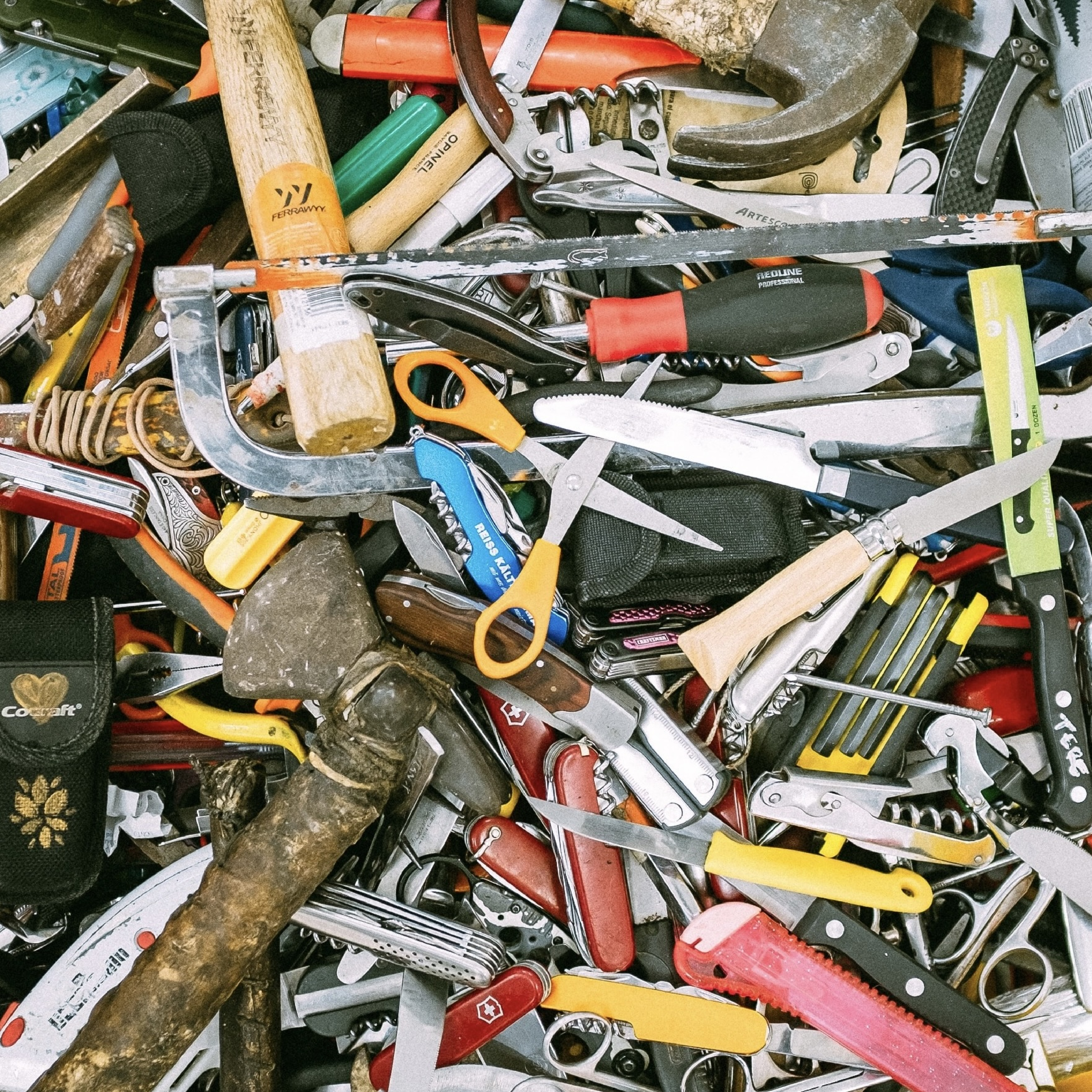}
            \caption{Before}
            \Description{Photograph of a cluttered drawer with various tools stacked, including hammers, saws, scissors, and knives.}
            \label{fig:image5}
        \end{subfigure}
        \hfill
        \begin{subfigure}{0.48\linewidth}
            \centering
            \includegraphics[width=\linewidth]{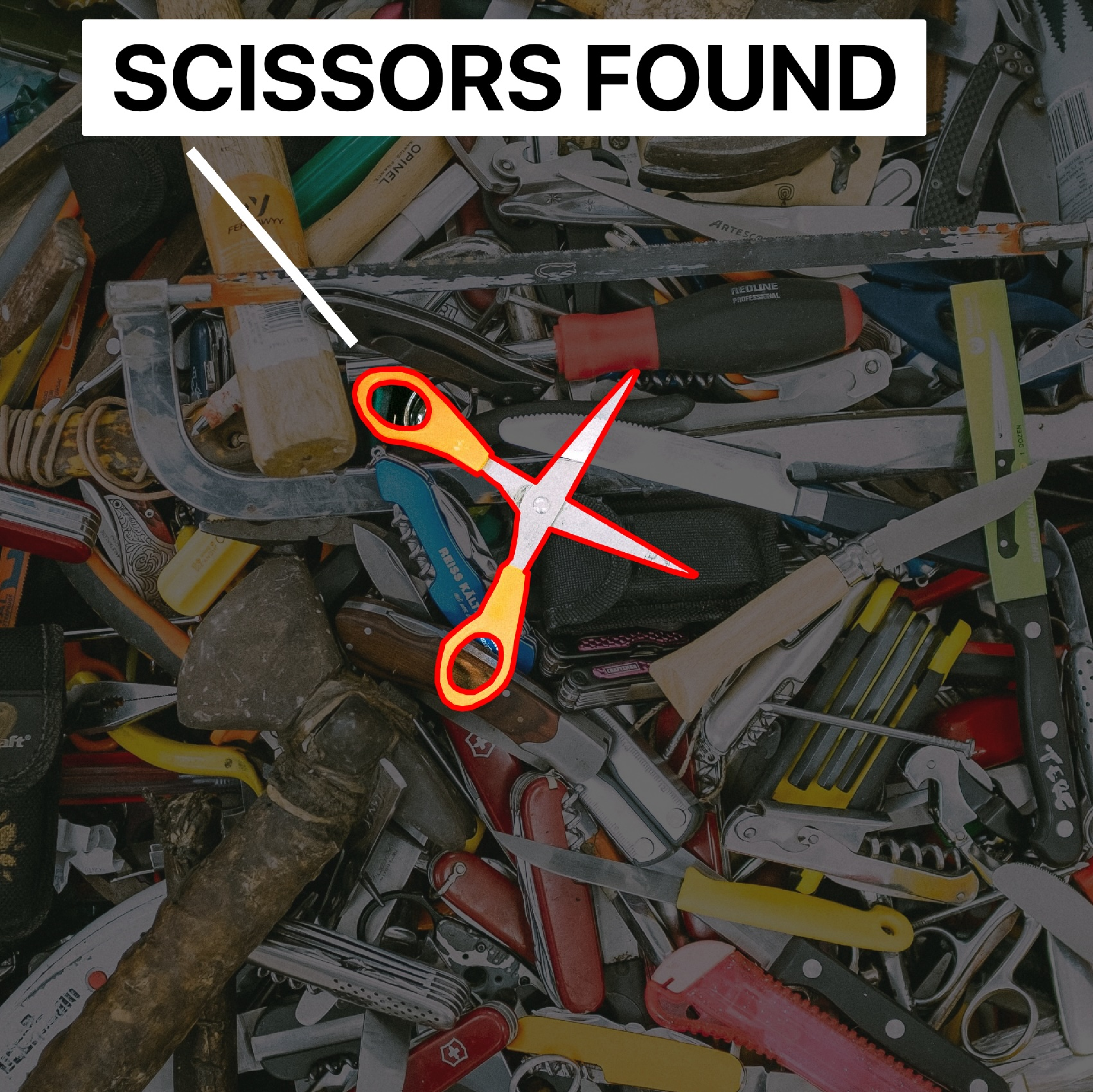}
            \caption{After}
            \Description{Same photograph as Figure 4E of a cluttered drawer. The pair of scissors is highlighted with a red border, while the rest of the image has a black tint to draw focus. Next to the scissors is a white rectangle with black text saying 'Scissors found,' connected by a line.}
            \label{fig:image6}
        \end{subfigure}
        \caption{Cluttered drawer, with a verbal request to locate `Scissors'. Device highlighting the scissors with cluttered background segmented. Original image: \cite{unsplash_tools}}
        \Description{Two subplots depict the before and after scenes of a cluttered drawer, showcasing how the device highlights a pair of scissors.}
    \end{subfigure}
    \hfill
    \begin{subfigure}{0.49\textwidth}
        \centering

        \begin{subfigure}{0.48\linewidth}
        \centering
        \includegraphics[width=\linewidth]{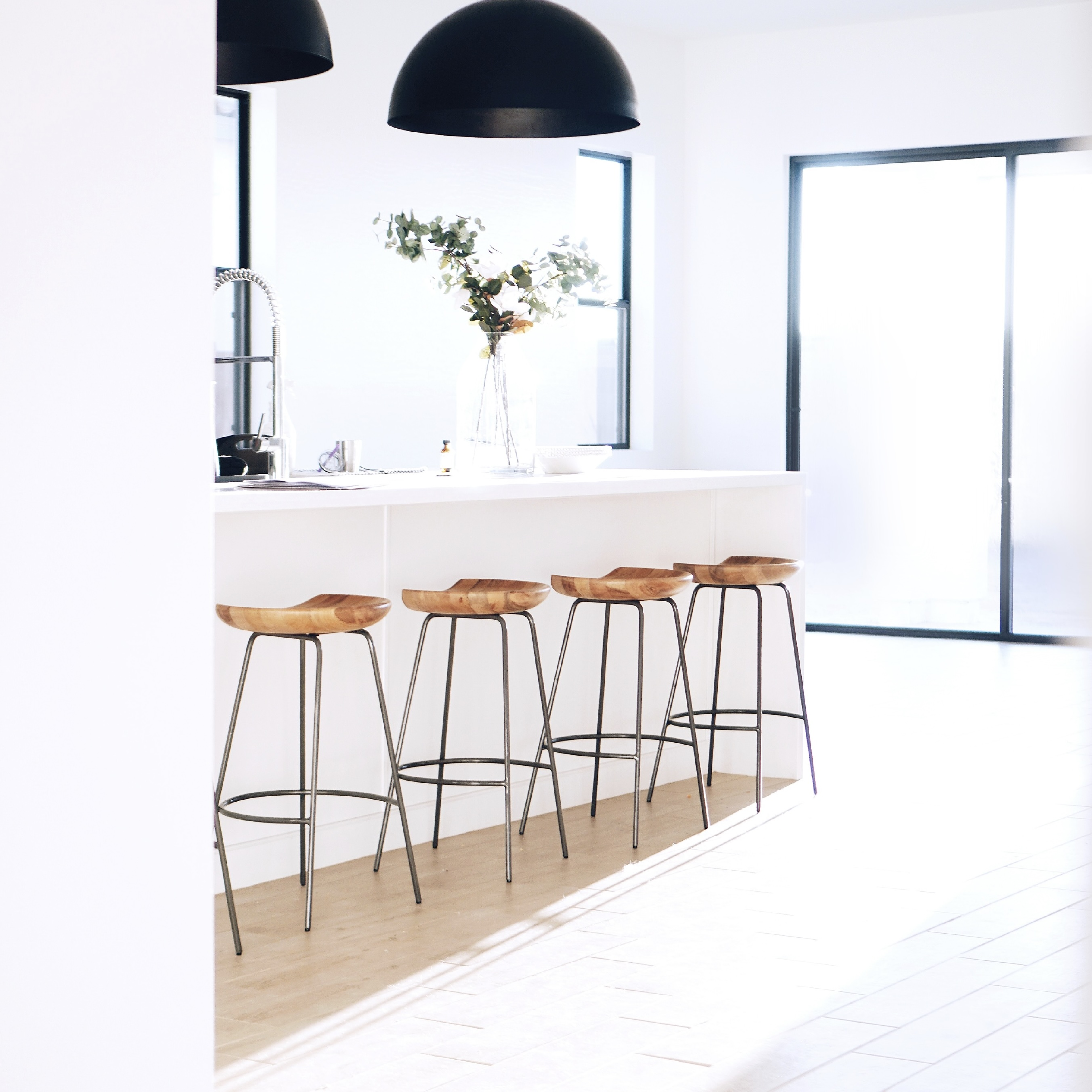}
        \caption{Before}
        \Description{Photograph of a white-coloured dining room with sunlight streaming in through a window, causing glare. A white kitchen counter with wooden stools is visible nearby. The white table and walls seamlessly blend together.}
        \label{fig:image7}
        \end{subfigure}
        \hfill
        \begin{subfigure}{0.48\linewidth}
        \centering
        \includegraphics[width=\linewidth]{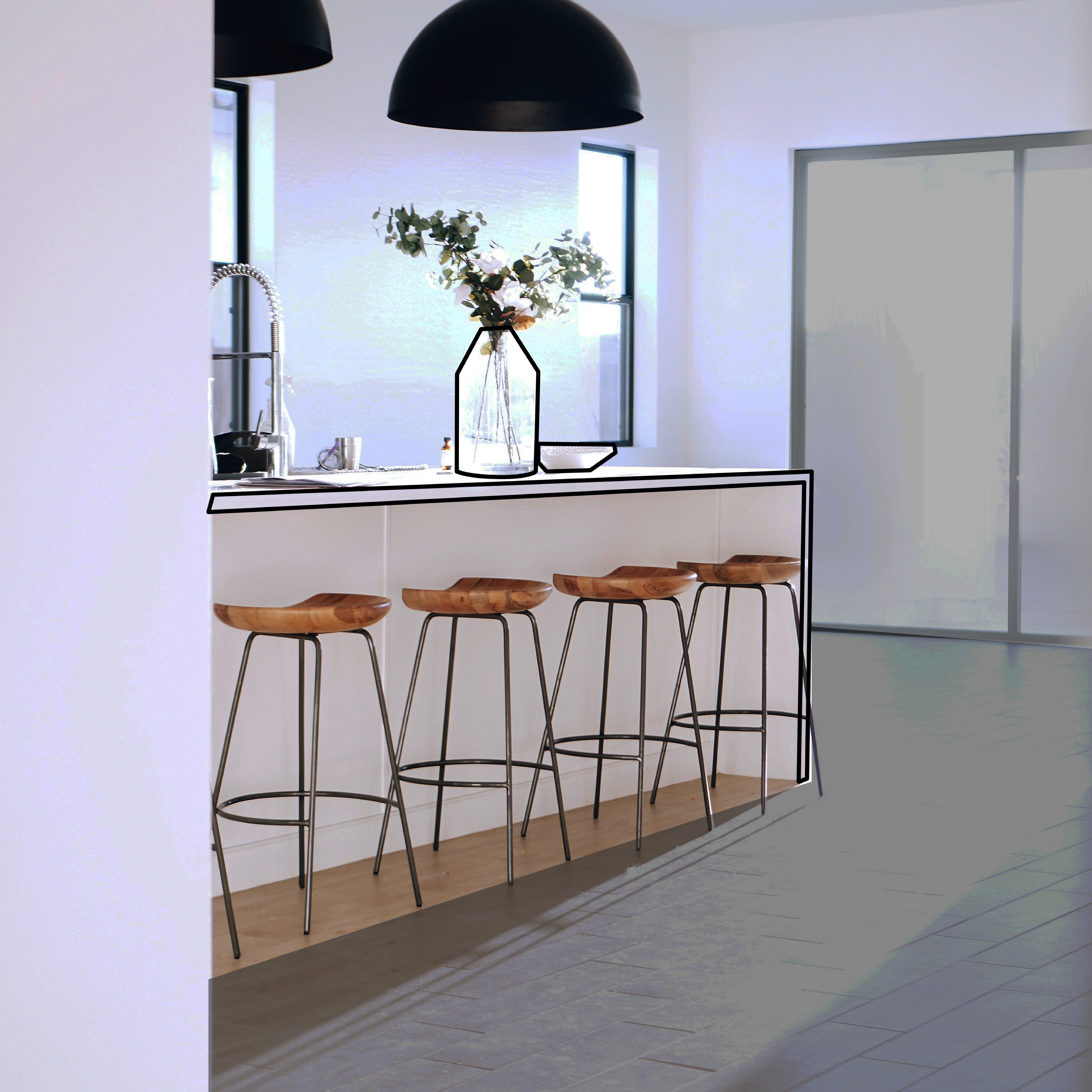}
        \caption{After}
        \Description{Same photograph as Figure 4H of a white-coloured dining room with a black tint applied over the window to reduce the bright light. A black border separates the white table from the walls.}
        \label{fig:image8}
        \end{subfigure}
        \caption{White coloured dining room with bright lights and sun glare. Device reduces glare, adjusting the bright area, and adding borders to the table for background separation. Original image: \cite{unsplash_glare}}
        \Description{Two subplots depict the before and after scenes of a white coloured dining room with bright lights and sun glare, showcasing how the device assists the user.}
    \end{subfigure}
    \caption{Example enhancements demonstrated to the participants during the focus group discussion.}
    \Description{Sequence of eight images in two rows demonstrating enhancements shown to participants.}
    \label{fig:demonstration-images}
\end{figure*}

To help participants understand VBAT devices, we utilised the term "smart glasses" in the study to denote devices that augment vision and participants were presented with images demonstrating the potential application of VBAT in various scenarios such as finding objects and people, glare reduction, and text enhancement. 
Figure \ref{fig:demonstration-images} illustrates samples of the images showcased during the demonstrations.
We then asked the participants about potential opportunities for VBAT to assist them. 
Given the unique requirement of each participant, we emphasised that we wanted to understand how it could assist them specifically.
A similar approach was employed by Ringland et al. \cite{ringland_asset2019} to explore the potential role of assistive technology in supporting psychosocial disabilities outside of a clinical or medical framework.

\subsubsection*{\textbf{Considerations when developing VBAT\@:}}
Lastly, participants were prompted to discuss potential barriers that could influence their decision to use or not use the device.

\subsection{Data Analysis}
Responses from the focus group discussion were transcribed and an inductive but deductively framed thematic analysis was adopted. 
Specifically, two researchers independently coded the transcripts to identify challenges, current strategies, opportunities, and potential solutions. 
After the initial coding, the two researchers reviewed, cross-checked, and revised the codes. 
Finally, all researchers collectively reviewed the codes to generate the set of challenges and device considerations.

\subsection{Challenges, Current Strategies, Opportunities and Potential Solutions}

Seven high-level challenges were identified from the focus group discussion (refer to Figure \ref{fig:challenges}).

\subsubsection*{\textbf{C1: Unawareness:}}
\label{sec: challenge1}
%Unawareness is defined as the challenge of lacking awareness of elements in one's surroundings, including objects, people, animals, shops, or text.
All seven participants reported challenges related to unawareness. 
Difficulties included being unaware of objects or people (n=6), experiencing Hemianopsia\footnote{Hemianopsia is visual loss in half of the visual field.}(n=4), being unaware of text (n=2), being unaware of moving objects-due to Akinetopsia\footnote{Akinetopsia is an inability to perceive motion, seeing the world as a series of static snapshots.}(n=2), and experiencing Hemineglect\footnote{Hemineglect is neglect of one side of the visual field, impacting awareness of surroundings.}(n=1).
Note that the difficulties discussed in the three focus groups were not cross-discussed; therefore, other participants may have encountered similar challenges but not raised them.

\subsubsection*{Current strategies:}
Two participants (P3, P4) use long canes to be alerted to obstacles in their immediate surroundings.
P1 reported that due to their lower visual field Hemianopsia, they use sensor lights at home to remind them of the staircase.

\subsubsection*{Opportunities and Potential Solutions:}
Five opportunities for VBAT devices in addressing unawareness were identified.
\begin{itemize}
\item Improving awareness of immediate hazards.
Discussion centered on the effectiveness of warning users about hazards by alerting them visually or simply highlighting potential hazards. 
Speed and reliability were pivotal, illustrated by examples such as alerting users of moving cars or reminding them to "mind the gap" when entering a train.
\item Improving awareness of objects, people, animals, and shops.
One proposed solution explored if VBAT devices could monitor and predict what the user might overlook based on their specific condition.
Then it can alert the user when something or someone enters their vicinity without the user being aware of it.
For alerting the user, one suggested solution involved changing to a bright colour such as orange, yellow, or red to draw attention to the object or situation.
Participants expressed that it could potentially help broaden their tunnel vision by increasing their overall awareness.
\item Improving awareness of text.
This could involve the device alerting the user to the presence of text, either by highlighting it or flashing a light to draw attention.
Another solution was to ensure text consistency, as variations in fonts, sizes, or colours can lead to text being overlooked.
P2 described their experience as follows:
\begin{quote}
\textit{It was instructions me daughter had left to look after me grandson, putting the most important thing in bright red across the top and I read this letter about 9 times to check I'd done everything, and I didn't see the red writing.}
\end{quote}
\item Raising awareness of the side affected by Hemianopsia. 
This could involve providing constant reminders or selectively alerting the user when something significant occurs on that side.
\item Modifying the visual field.
An example of this was bringing areas of visual neglect into the conscious visual field. 
However, unlike an optical solution such as a prism lens \cite{giorgi2009clinical, peli2000treating}, which would always alter their vision when worn, participants discussed the possibility of selective modification.
\end{itemize}

\subsubsection*{\textbf{C2: Locating:}}
Locating refers to the difficulty of finding an object that you know is in your environment.
%being aware that something is present but not being able to determine its exact location.
Four participants reported challenges with locating and as P2 described it:
\begin{quote}
\textit{You cannot find what you're looking for, even though it's right in front of you.}
\end{quote}
Examples of difficulties included locating objects (n=2), spatially missing objects (n=2), difficulties in locating due to the inability to control visual fixation (n=2), and locating people (n=1). 
P1 described their experience with controlling visual fixation:
\begin{quote}
\textit{I'm assuming a normal person with normal sight that has no visual impairment, it comes automatic to lock on and lock off, to fixate and unfixate on things, whereas with someone with CVI, it's a real struggle to lock on to something.}
\end{quote}
One of the underlying issues contributing to the difficulty of locating items was visual clutter (n=3). 
This was highlighted by RP:
\begin{quote}
\textit{One of the biggest things for me [...] is being able to find an item, a specific item, in a cluttered background. [For example]
going to the supermarket and trying to find a can of peaches amongst all the cans of fruit.}
\end{quote}

\subsubsection*{Current strategies:}
P2 described one strategy of painting/labelling objects around the house, such as the phone, TV remote, and garden tools, in a bright orange colour.

\subsubsection*{Opportunities and Potential Solutions:}
One of the primary opportunity for VBAT devices lies in assisting users with locating objects, people, shops, animals or text. 
Various solutions were discussed, with the most suggested approach involving highlighting the object when the user verbally requests it. 
This highlighting could be done by flashing a light, changing the target to a bright colour, adding boundaries around the target, or directing the user with arrows or pointers. Additionally, P4 and P6 suggested adding directional audio directions, instead of visual cues.
% VBAT also needs to effectively identify items within cluttered environments, as articulated by P3: \textit{"Something that could pick out through the clutter and find stuff"}.
Participants also discussed the potential to locate objects 360-degrees around them, as they may not always be able to orient the device cameras in the correct direction.

\subsubsection*{\textbf{C3: Identifying:}}
%After becoming aware of and locating an object, the next challenge was identifying it.
All seven participants reported difficulties in identification: people (n=4), objects (n=1), animals (n=1), facial expressions (n=1).
Four participants reported simultanagnosia, that makes identification challenging. 
As P1 described:
\begin{quote}
\textit{Do you see the forest from the trees, or do you see the whole forest? For me, I'd see the forest from the trees. I don't see the whole forest.}
\end{quote}

\subsubsection*{Current strategies:}
P2 employed an approach where they look around the face to refresh the image when attempting to identify a person. 
This strategy is also known as the Wagon Wheel approach \cite{Scotland_2021}.
RP similarly employs this method to comprehend visual scenes, stating: 
\begin{quote}
\textit{It gives me the ability to build up the visual scenes slowly, one item at a time, so it does not overwhelm my brain.}
\end{quote}

\subsubsection*{Opportunities and Potential Solutions:}
Participants discussed the potential of VBAT devices to assist in the identification of people, objects, and animals.
For people, one solution discussed was highlighting the face and displaying the identified person's name. 
Also, enhancing facial features such as adding borders or increasing sharpness could make features more prominent for identification.
For objects, reducing the size of larger objects to fit within the smaller visual attention field could potentially aid in seeing the bigger picture.
For animal identification, both highlighting and auditory identification methods were suggested.

\subsubsection*{\textbf{C4: Reading:}}
After becoming aware of text and locating it, the subsequent challenge is reading it.
Six participants reported difficulties in reading text in their surroundings, attributed to factors such as colour, font, and size.
P1 described this:
\begin{quote}
\textit{If it's just plain, with no accentuations on each individual letter, I'll be able to see. [...] I'll struggle with anything that's a little bit outside of this simpler sort of format [Times New Roman font].}
\end{quote}

\subsubsection*{Current strategies:}
Participants discussed current strategies, including using only plain fonts (P1), using the Mini Ruby CCTV to convert text to a white background and black text (P4), restricting reading to text on the computer screen (P5), and preferring audio-books over reading books (P3). 
Many participants try to avoid reading altogether, as noted by P5:
\begin{quote}
\textit{I don't read books or newspapers or magazines anymore. It's just too hard.}
\end{quote}

\subsubsection*{Opportunities and Potential Solutions:}
Participants discussed ways to enhance text legibility for improved readability.
These methods included standardising the text by adjusting the font, font size, and font colour or altering the text to feature a black font on a light background.
%P1 also noted that fonts with serifs, such as Times New Roman, could make reading more challenging, suggesting that the device should prioritise converting to san-serif fonts like Arial.
Lastly, in addition to visual enhancements, participants discussed the option of having the device read the text aloud.

\subsubsection*{\textbf{C5: Sensory overload:}}
All seven participants highlighted the challenge of sensory overload, which can arise from being inundated with multiple forms of sensory inputs.
This can lead to feelings of stress and sometimes induce a tunnel vision-like effect as the brain struggles to process the overwhelming stimuli. 
Additionally, participants noted that sensory overload may result in difficulties maintaining visual attention. 
One example discussed by two participants was engaging in face-to-face conversations. 
P1 articulated this difficulty:
\begin{quote}
\textit{Having a conversation and locking eyes and being able to filter out the background noise, that's normal. But for someone with CVI, that's exceptionally hard, and next to impossible to do it.}
\end{quote}

\subsubsection*{Current strategies:}
Participants shared current strategies for managing sensory overload by maintaining visual attention.
Three participants (P2, P4, P6) mentioned consciously making an effort to maintain visual attention, particularly when walking or driving. 
During face to face conversations, P2 reported blurring their vision, while another stated they look away to reduce sensory overload.
Additional strategies include using voice control on smartphones to avoid looking at the screen (P5), and utilising browser extensions to declutter content when using their computer (P4).
P1 described how they discovered that wearing contact lenses helped them maintain visual attention as a side-effect of them narrowing the peripheral vision.
Lastly, P3 and P4 discussed actively managing their visual stamina but monitoring their energy levels and adjust their activities accordingly. 
P4 explained:
\begin{quote}
\textit{I'm good at recognising how tired my brain is, [...] I was falling so much before I would realise, Oh! you must have been tired. So, I know, zero and 100, and I'm not very good at knowing the numbers between that.}
\end{quote}

\subsubsection*{Opportunities and Potential Solutions:}
The opportunities for VBAT devices were explored along two distinct avenues:
\begin{itemize}
    \item Reducing sensory overload. Participants proposed filtering out irrelevant information using the device, such as blurring or dimming unnecessary objects. Additionally, actively removing background noise during face-to-face conversations was suggested. 
    \item Aiding visual attention. Participants also discussed the potential for the glasses to detect signs of sensory overload and assist in regaining visual attention. Methods for detecting anxious states included monitoring eye movement or heart rate. Participants also considered methods such as reducing peripheral vision akin to contact lenses and guiding the user's visual attention by highlighting or displaying a target area to focus on.
\end{itemize}

\subsubsection*{\textbf{C6: Mobility:}}
Participants discussed challenges in mobility stemming from two difficulties - visuomotor coordination difficulties (n=2) and difficulty navigating unfamiliar environments (n=2).
P3 elaborated:
\begin{quote}
\textit{The first time I go into an [unfamiliar] environment, I have to slow way down and I do behave as if I'm totally visually impaired. I mean I am visually impaired, but like I do all the things that a totally blind person would do to figure out the space and then the next time I'm there my brain is mapped for it.}
\end{quote}

\subsubsection*{Current strategies:}
Participants mentioned strategies such as memorisation (P1, P3), planning routes in advance (P3, P4), relying on orientation and mobility training (P3), and using a walker to aid visuomotor coordination (P6).

\subsubsection*{Opportunities and Potential Solutions:}
Participants discussed opportunities for VBAT devices to assist with route finding by visually outlining the route or providing auditory instructions. 
%It's important to note that 
The opportunities and solutions discussed in Challenge 1: Unawareness (Section \ref{sec: challenge1}) can also be applicable when navigating unfamiliar environments.

\subsubsection*{\textbf{C7: Luminance and Contrast Sensitivity:}}
All seven participants reported difficulties concerning luminance and contrast sensitivity.
Four participants specifically discussed challenges with subject and background separation, particularly when the colours are similar. 
P2 described it:
\begin{quote}
\textit{So if there's a colour behind, it'll sort of blend in with the face, so the whole thing is quite complicated.}
\end{quote}
Additionally, four participants reported sunlight and bright lights causing glare as a difficulty.
Other reported difficulties included light streams at night (n=2), dim lighting causing issues (n=2), experiencing visual snow\footnote{Visual snow is a persistent, dynamic visual disturbance, resembling static, flickering lights, or sparkling dots in vision.} (n=1), and difficulty perceiving muted colours (n=1).
This, in turn, can lead to or exacerbate other challenges such as identification and reading.

\subsubsection*{Current strategies:}
Three participants (P1, P4, P5) reported improving lighting at home to aid in subject and background separation.
Other approaches included employing bright colours, such as orange, for various objects (P2), and dressing their dog in a bright-coloured t-shirt (P3).
When outdoors, P3 and P4 mentioned relying on changing cues in lighting and shadows, particularly on the pavement, to assist them.
To mitigate glare, participants mentioned wearing sunglasses (P2, P3) and a low-brimmed hat (P2).
For dealing with visual snow, P3 mentioned using a FL41 tint lens \cite{reyes2024fl}.
Regarding current strategies with light streams, P3 stated they completely avoid driving at night.
As described by P3:
\begin{quote}
\textit{I won't drive at night. I mean, in fact, I keep track of when the sun sets and I make sure that I'm on my way and at home.}
\end{quote}

\subsubsection*{Opportunities and Potential Solutions:}
\begin{figure}[H]
    \centering
    \begin{subfigure}{0.49\textwidth}
        \centering
        \begin{subfigure}{0.49\linewidth}
        \centering
        \includegraphics[width=\linewidth]{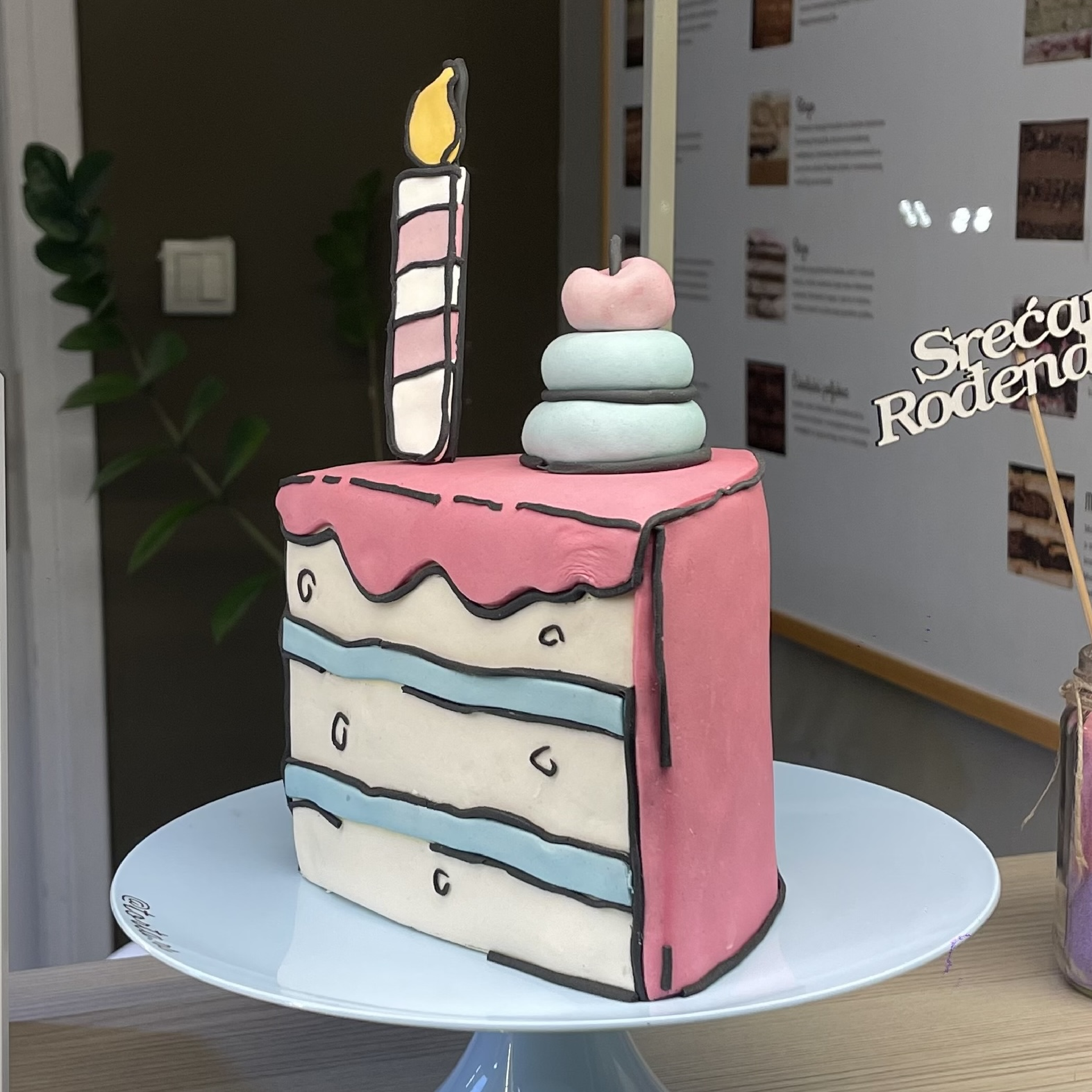}
        \end{subfigure}
        \begin{subfigure}{0.49\linewidth}
        \centering
        \includegraphics[width=\linewidth]{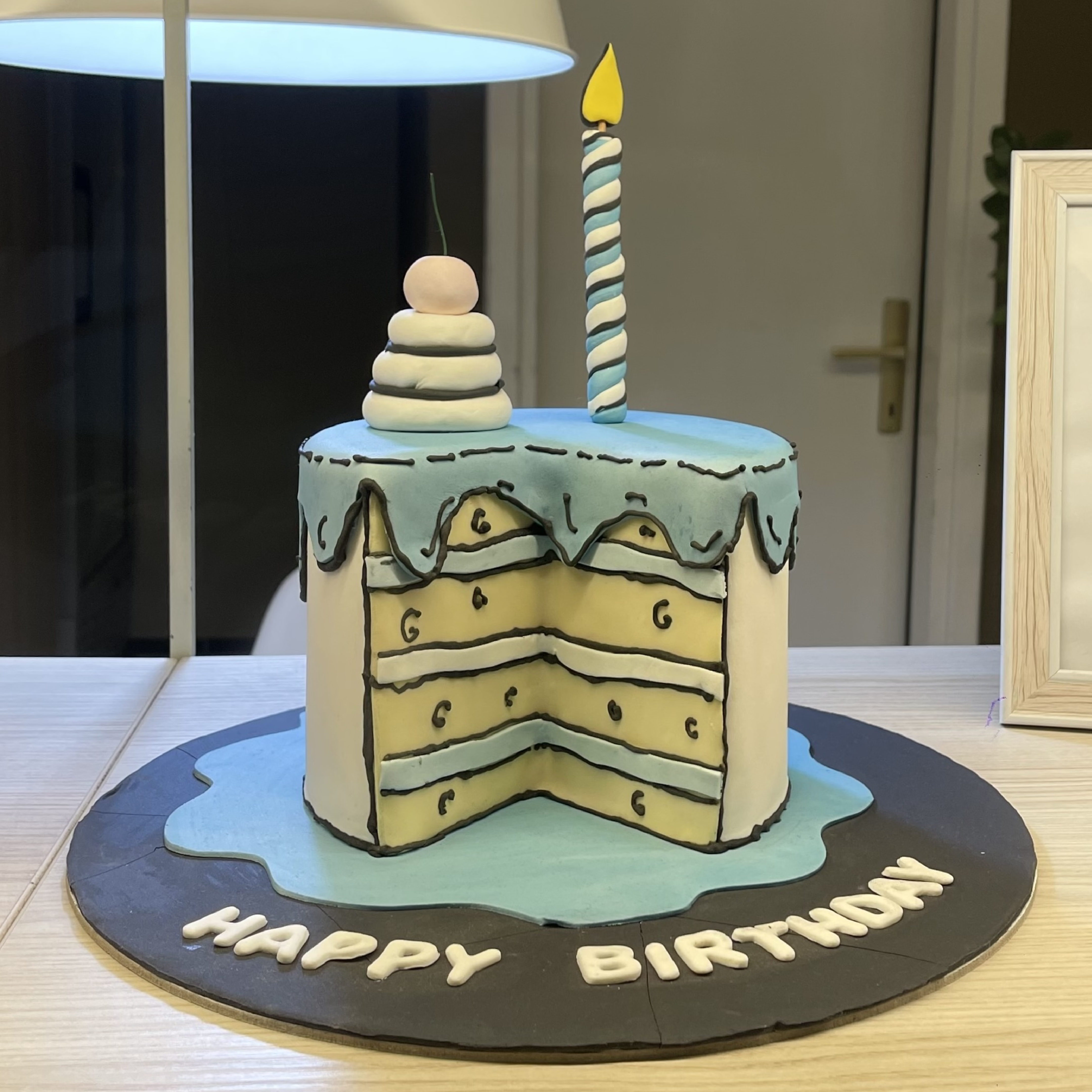}
        \end{subfigure}
    \end{subfigure}
    \caption{Image shared by P1 demonstrating the use of boundaries around objects to aid in subject and background separation.}
    \Description{Two pictures of birthday cakes, one showing a slice and the other depicting a cake with one slice removed. Both cakes have a black icing border around the edges, icing and filling to accentuate their features.}    
    \label{fig:cakes}
\end{figure}

Two main opportunities for VBAT were discussed:
\begin{itemize}
    \item Adjusting the brightness of the scene. This includes solutions such as artificially reducing bright lights, cutting out glare and brightening up the scene.
    \item Improving background and subject separation. This could be done by adding boundaries around objects. Participants discussed that the device could be smart and only add a border when it detects a similar-coloured backgrounds behind the subject (person or object). Figure \ref{fig:cakes} displays a photograph observed by P1 at a cake shop, depicting the use of borders around objects. Other participants confirmed that such borders were beneficial to them as well.
\end{itemize}

\subsubsection*{\textbf{Dependence on Others:}}
Five participants (P1, P2, P4, P5, P6) noted that these challenges led to a lack of independence and reliance on other individuals, including family, friends, and aids. 
While dependence on others is not inherently a difficulty or challenge, it becomes a factor due to the various challenges they face, preventing them from achieving full independence.
P5 shared their experience of relying on other people due to difficulties with unawareness:
\begin{quote}
\textit{Well if I go anywhere like go to a hospital for an appointment, I've got to have somebody with me because if I wander around the corridors, unless I'm constantly scanning I don't see all the turns.}
\end{quote}

\subsection{VBAT Device Considerations}
Nine considerations emerged from the focus group discussions, identified as crucial for researchers to consider during the development of VBAT devices.

\subsubsection*{\textbf{Adaptability to user needs and preferences} (P3, P4, P5, P6, RP\@):}
P3 highlighted the diversity and uniqueness of requirements among people with CVI, stating, \textit{"No two CVI %[is that] 
[...] are the same."} 
Various factors, such as the onset of the CVI condition (e.g., stroke, acquired brain injury, in utero), age of onset (birth, young, adult), and duration of the condition, can influence the level of assistance required from these devices. 
As P3 further explained:
\begin{quote}
\textit{There is a big difference in how long you live with it, and how are you at adapting and coming up with strategies that you can repeat in the moment.}
\end{quote}
Therefore, customisation for each person's condition is essential. 
Examples include compatibility with existing glasses, support for multiple forms of input and accommodating speech impediments for voice input. 

Additionally, the device should augment the user's abilities rather than replace them. 
For instance, if the user already uses a long cane, the device should complement its functionality rather than duplicating the information provided by the cane.

\subsubsection*{\textbf{Device's impact on sensory overload} (P1, P2, P3, P5, RP\@):}
Participants voiced concerns about the device exacerbating their stress and further leading to sensory overload. 

\subsubsection*{\textbf{Device ergonomics} (P1, P2, P3, P4, P5, P6, RP\@):}
All participants expressed concerns regarding device ergonomics, emphasising the importance of comfort, lightweight design, and extended battery life. 
Additionally, it needs to be discreet and fashionable.

\subsubsection*{\textbf{Ease of setup} (P5):}
The setup process for the device should be straightforward and seamless. 
When inputting personal details such as family members and personal items (wallet, phone, or car), the device should be able to learn quickly.

\subsubsection*{\textbf{Processing speed} (P3):}
The speed of processing is crucial, especially to alert them of real-time hazards. 
As P3 questioned:
\begin{quote}
\textit{I kind of would wonder how this type of a tool would be able to intervene quick enough in real time to catch that kind of mistake [falling down a staircase]?}
\end{quote}
\vspace{1mm}
\subsubsection*{\textbf{Accuracy of device} (P5):}
The accuracy and reliability is paramount for gaining user trust and reducing confusion. 
P5 shared their concern about accuracy of reading text:
\begin{quote}
\textit{As long as it reads it in the right way. I have tried an app on the phone that was meant to read text to you but when you show it like a newsletter page, and it's got columns, it just reads across the column. It just doesn't make any sense. Because it doesn't have this smarts to say I've reached that side of the column and need to go back to the next line. It just reads across the whole page at once.}
\end{quote}
\vspace{1mm}
\subsubsection*{\textbf{Privacy} (P1, P5, RP\@):}
Three participants expressed concerns regarding the privacy implications of the devices, particularly when using voice commands in public and employing face recognition technology in public settings. 
Another concern was the handling of sensitive data by the device and ensuring that such information is not audibly shared in public.

\subsubsection*{\textbf{Use of additional sensors} (P2, P5, RP\@):}
Finally, participants explored the integration of other sensors or technologies such as heart-rate monitoring, eye-tracking to detect if the user is experiencing sensory overload, and GPS to assist with outdoor navigation.

With the exception of the `Device's impact on sensory overload' for people with CVI and use of additional sensors to detect sensory overload, other design considerations align with those common in accessibility research.

\section{Discussion}

This section discusses the findings from the two studies and the implications for AT device research for people with CVI.

\subsection{What are the Similarities and Differences in AT Needs Between People with CVI and OVI?}

It is tempting to assume that research designed for OVI could be readily repurposed to benefit those with CVI.
This is especially true for Ocular Low Vision (OLV) studies \cite{zhao2015foresee, lang2020augmented, zhao2016cuesee}, as they also provide visual modality in addition to audio and haptic.
However, based on the findings from our study and prior literature on CVI, we have synthesised five main differences between the needs of people with CVI and OVI that must be considered when designing AT:

\subsubsection*{\textbf{Functional Vision Difficulties:}}
We first need to consider the similarities and difference between the two levels of functional vision difficulties within the two groups.
As outlined in Section \ref{sec:low-high-function}, people with ocular blindness and OLV face challenges primarily in low-level functional vision, whereas those with CVI may experience difficulties mainly in high-level functional vision.

The distinction becomes more complex because people with CVI can also experience low-level functional difficulties. 
This was evident in the focus group discussions when participants mentioned challenges related to visual field such as Hemianopsia and Hemineglect. 
These conditions can manifest in both OLV and CVI and are determined by the location of damage (eye or brain) in the visual processing system. 
Therefore, for individuals facing difficulties in low-level functional vision, devices and applications tailored to OLV may indeed be applicable.

However, in addition to low-level functional difficulties all participants in this study experienced high-level functional difficulties.
As P4 expressed:
\begin{quote}
\textit{I don't understand where one of my vision thing ends and the other begins.}
\end{quote}
Evidence from the educational field indicates that strategies for people with OVI may not be effective for people with high-level functional vision difficulties \cite{martin2016cerebral}.
During the focus group discussion, P1 posed a question to P2 that illustrated this point:
\begin{quote}
\vspace{1mm}
\textit{P1: So when that writing appears [someone making them aware of text], and it's like magic, are you then able to understand the meaning of what's being presented in writing?}
\end{quote}
\begin{quote}
\textit{P2: Yeah, totally, I see it, and I understand it.}
\end{quote}
\vspace{1mm}
It is apparent that P2's struggle with finding text does not stem from issues with low-level functional vision; instead, the difficulty lies in their awareness and ability to locate text, which pertains to high-level functional vision.
Therefore, people with CVI may face similar challenges to those with OVI, but combined with their high-level functional difficulties, the underlying difficulties could be quite different, necessitating careful consideration to ensure that AT effectively addresses their specific needs.

\subsubsection*{\textbf{Single vs Multi-Modality Interaction:}}

Sighted humans acquire approximately 80\% of environmental information through vision \cite{man2018possibilities, BrainLine_2018}. 
As studies have pointed out, people with CVI with high-level functional vision difficulties predominantly rely on their vision for daily activities \cite{chandna2021higher}. 
This was evident among all seven participants, who primarily used vision to comprehend their surroundings. 
P5 explained:
\begin{quote}
\textit{It's just that basically human beings get most of their information in visually.}
\end{quote}
However, the primary distinction between people with OLV and CVI is that those with CVI have consistently shown that simultaneous integration of multiple modalities—such as visual, auditory, and haptic—can overwhelm them \cite{philip2014identifying, lam2010cerebral}. 
We observed this with all seven participants, and it was clearly articulated by P1:
\begin{quote}
\textit{That would be probably too much information overload. [...] Auditory and visual problems coming at you once, and your brain is unable to filter it out, so that would be just doing the opposite of what you are asking me to do.}
\end{quote}
This is in contrast to people with OLV who have shown preferences combining both audio and visual modalities during way-finding \cite{zhao2020effectiveness}.
This also means that people with CVI necessitate interactions that are less mentally taxing and overwhelming compared to neurotypical individuals with OLV. 
Consequently, when developing VBAT devices, it is crucial to prioritise vision as the primary modality and avoid overusing audio and haptic modalities that may overwhelm the user.

\subsubsection*{\textbf{Complexity Impact on Vision:}}
Previous studies have shown that people with OLV maintain relatively stable performance whereas people with CVI experience escalating difficulties as visual complexity increases \cite{bennett2018assessing}. 
This was evident in our focus study, particularly discussed in the challenge of locating.
Our study revealed insights from two participants suggesting that visual stamina could be a critical factor as well.
With heightened visual complexity, visual stamina diminishes, resulting in vision-related challenges.
Hence, the manner in which devices visually interact with users becomes pivotal, especially in tasks involving high complexity. 
P3 elaborated on this aspect:
\begin{quote}
\textit{So we have conversations about CVI meltdowns, and [...] how do you get it [VBAT device] so this isn't just one more piece of information at a time on the brains already overloaded and beyond its bandwidth. [...] and the brain shuts down, and the next thing they know they're tripping and falling because it was too much and it wasn't a behavioural miscue, it's just that the brain is overloaded.}
\end{quote}

\subsubsection*{\textbf{Association with Other Neurological Conditions:}}
CVI commonly co-occurs with other neurological conditions such as cerebral palsy (CP) \cite{lueck2015vision}, which has been shown to indirectly impact vision \cite{martin2016cerebral}.
This is in contrast to those with OLV, which primarily affects the eyes and can often be isolated from neurological conditions. 
We noticed the impact on vision during the focus group discussion with the two participants who had CP.
Both of them showed a preference for audio-based assistance, despite relying primarily on vision for perceiving their environment. 
This preference aligns with findings from Dutton et al.'s work, indicating that children with CP may prioritise auditory input due to the cognitive demands associated with controlling physical movements and posture, leaving limited mental capacity for visual engagement \cite{lueck2015vision}.
Another study also reported that children with both CP and CVI face increased physical and functional challenges compared to those with CP alone \cite{salavati2014gross}, suggesting an indirect impact on their vision. 
This observation is also consistent with the finding that people with CVI prefer a single modality discussed above.
However, for future studies, we suggest that researchers report and discuss other neurological conditions experienced by participants.

\subsubsection*{\textbf{Effects of Vision Rehabilitation:}}
Unlike OVI, CVI is a brain-based dysfunction, allowing assistive devices to facilitate permanent alterations in the brain's visual pathways through neuroplasticity \cite{bennett2020neuroplasticity}. 
Numerous reviews and studies have explored rehabilitation techniques and methods leveraging neuroplasticity and brain modification to enhance vision \cite{delay2023interventions, mcdowell2023review, weden2022evidence, waddington2017review, ciman2013helpme}. 
However, the majority of these studies focus on the paediatric or young adult populations, similar to what we observed in our scoping review, necessitating further research to understand their effects on adults.

Numerous studies have investigated the potential of OLV rehabilitation \cite{stelmack2001quality, agarwal2021current}, including one study identified in our scoping review \cite{lorenzini2021personalized}, yet the extent of improvement has been observed to be relatively moderate \cite{lamoureux2007effectiveness}. 
In contrast, people with CVI maintain functionality in eye structures like the retina and optic nerve. 
This enables assistive devices to potentially modify neuropathways over time to utilise these and thus enhancing their vision.
P3 expressed optimism about this prospect:
\begin{quote}
\textit{The power of introducing this to a child, at the age of like four or five, all of a sudden a lot of the miscues that define their life gets redirected in a very positive way so they're not constantly colliding with their space, or the people in it.}
\end{quote}
However, it is essential to recognise that this effect may have drawbacks, potentially resulting in lasting vision changes due to reliance on assistive devices. 
Therefore, researchers investigating assistive devices for CVI should exercise caution and consider long-term effects through extended studies.

\subsubsection*{\textbf{Summary and Implications for Designing Assistive Technologies\@:}}
%Considering the unique challenges faced by people with CVI, we can identify distinct differences in their AT needs by analysing the specific difficulties and current strategies employed to address them.
Table \ref{tab:differences} provides a concise summary of the key similarities and distinctions among people who have ocular blindness, OLV and CVI. 
Understanding these nuanced differences is crucial for developing effective assistive technologies.

Among the challenges, ‘C5: Sensory Overload’ stands out as the challenge most unique to people with CVI. 
However, while the other challenges are also observed in people with OLV\cite{szpiro2016finding, szpiro2016people},  they occur for different reasons. 
These differences impact on how AT can be used to address these challenges.

For instance, in the case of `C2: Locating,’ people with CVI experience difficulties due to visual clutter around objects, while people with OLV struggle because they cannot see the object clearly.
Similarly, ‘C3: Identifying’ highlights another significant difference. 
People with OLV cannot identify people because their eyes do not see the person clearly, whereas those with CVI may see the person but have difficulty because the part of the visual system that connects visual information to identification does not function properly. Therefore, enhancing visual clarity by, say, magnification ~\cite{ar_magnification} could help people with OLV but not those with CVI, who may be better helped by a name next to the person to aid in identification.
Therefore, while there are overlapping challenges between CVI and OLV, each condition requires tailored strategies to address the unique aspects of their visual impairments. 

\begin{table*}[t]
\centering
\caption{Similarities and Differences among people who are blind, have low vision, and have CVI}
\begin{tabular}{|p{3.6cm}|p{3cm}|p{4.3cm}|p{5.5cm}|}
\hline
\textbf{Criteria} & \textbf{Blindness} & \textbf{Low Vision}   & \textbf{CVI}          \\ \hline
\textbf{Functional Vision \newline Difficulties}   & Low-Level                        & Low-Level             & Low-Level \& High-Level            \\ \hline
\textbf{Modalities Preference}       & Multiple Modalities               & Multiple Modalities & Single Modality \\ \hline
\textbf{Complexity Impact on \newline Vision}   & None                        & No Impact             & High Impact            \\ \hline
\textbf{Association with Other Neurological Conditions}   & Sometimes  & Sometimes             & Frequently             \\ \hline
\textbf{Effects of Vision \newline Rehabilitation}   & Little Impact  & Little Impact     & High Impact            \\ \hline
\textbf{Visual Difficulties \newline Examples}   & Minimal to no visual perception                        &  Difficulty seeing objects, sensitive to contrast, identifying colour  & Finding objects in clutter, unawareness of text, finding a relative in a crowded room            \\ \hline
\end{tabular}
\label{tab:differences}
\end{table*}

\subsection{How Practical are the Solutions that were Discussed?}
Given the limited work focused on CVI, this section focuses on work from other fields and areas to determine the state-of-the-art (SOTA) and practicality of the solutions discussed during the focus group discussion.
The technologies for the potential solutions were separated into three main categories based on the sub problem they address:

\subsubsection*{\textbf{Contextual understanding:}}
Contextual understanding refers to the capability of a device to perceive and comprehend its surroundings based on the inputs it receives. 
These inputs can come from cameras, microphones, heart rate sensors and eye tracking devices. 
The majority of algorithms capable of interpreting these inputs rely on some form of machine learning. 
Proposed solutions for contextual understanding can be categorised into understanding the environment and understanding the user.

\textit{Understanding the environment} is fundamental to address all challenges. Accurately reconstructing the user's immediate surroundings enables the device to effectively assist the user with various challenges. This can be addressed through two main avenues: 
\begin{itemize}
    
    \item Identify objects and people. Numerous studies in computer vision have focused on tasks ranging from object detection \cite{ramik2014machine, jiao2019survey} to object tracking \cite{zhang2021recent, elhoseny2020multi}, enabling the device to locate and continuously monitor objects and people. 
    However, as highlighted by P5, the ease of setup is paramount. 
    Therefore, studies focusing on single-image face recognition \cite{ding2016comprehensive, tan2006face, wu2002face} are relevant for training the device with minimal images.
    \item Assist with route finding. Multiple studies cover various aspects of indoor route finding \cite{huang2010survey, liu2011door} and outdoor navigation \cite{el2023survey, alwi2013survey}. Numerous commercial solutions such as HyperAR \cite{hyper_ar}, Google AR maps \cite{google_ar}, and Apple AR maps \cite{apple_ar} further support this field.
    
\end{itemize}

\textit{Dynamic understanding of the user} is pivotal to address C1, C2 and C5, particularly considering the variability of CVI among individuals and the necessity for the device to adapt to user preferences and requirements. This can be addressed through two main avenues:
\begin{itemize}
    \item Predict user's visual blind spots.
    Being able to understand and predict what the user is perceiving allows the devices to selectively intervene and thus reduce unnecessary information.
    Work in computer vision, particularly in autonomous driving and driver assistance systems, has investigated algorithms to predict the driver's awareness of objects or people in their surroundings \cite{gandhi2006pedestrian, phan2014estimating, carnie2005computer, sharma2016practical, gupta2021computer}. 
    
    \item Anticipate sensory overload.
    As participants proposed, heart rate and eye tracking can be used to detect sensory overload states.
    Studies in biomedical engineering have explored machine learning approaches to predict stress states \cite{haque2024state, albaladejo2023evaluating, banerjee2023heart, mlmentallstress} and sensory overload \cite{amadori2021predicting, deng2022sensor, ponce2024interpretable, truaistar2023sensory} using heart rate sensors.
\end{itemize}

\subsubsection*{Practicality:}
Currently, machine learning algorithms encounter significant challenges in reliability. 
There are several causes for this, but for the solutions discussed in the study, we identified two main ones.
\begin{itemize}
    \item Real word conditions. Discrepancies often arise between their reported accuracy and real-world performance due to differences between the ideal conditions and actual operating environments. 
    This is caused by factors like varying lighting conditions, object sizes, and image angles. 
    Participants stressed the importance of accuracy in the device consideration, as people's trust in these algorithms is contingent on observed accuracy \cite{ml_accuracy}.

    \item Constraints on low-power consumer devices. 
    For example, while SOTA object detection algorithms can achieve an accuracy of 85\% \cite{object_detection_review}, they are typically evaluated on high-performance hardware with large models containing thousands of parameters. 
    Additionally, performing multiple object tracking requires significant computational resources. 
    Consequently, there exists a constraint in achieving both speed and accuracy on edge devices. 
    As highlighted by P3, in critical applications like hazard detection, it is crucial for these algorithms to maintain high levels of accuracy while also operating at fast speeds.
\end{itemize}

Therefore, further advancements are needed in this domain for contextual understanding to become practical in real-world settings. 
Two potential approaches that AT and AI researchers can adopt to help address these challenges are:
\begin{itemize}
    \item Datasets for contextual understanding. 
    A significant limitation of much of the machine learning and deep learning research lies in its heavy reliance on available datasets for training and testing algorithms. 
    Datasets play a pivotal role in both developing novel machine learning algorithms and fine-tuning existing models to support people with CVI in AT fields.
    The VizWiz dataset \cite{gurari2018vizwiz} serves as a prime example of an object detection dataset by people with vision impairment, featuring sub-datasets addressing diverse issues such as image quality and visual privacy. 
    Thus, to fully leverage the potential of machine learning algorithms, researchers in AT should work together with people with CVI to publish openly available datasets.

    \item Investigating user interactions for model uncertainty.
    Given the SOTA object detection algorithms only achieve 85\% in ideal conditions, it is expected to be uncertain.
    This is similar to the human vision, which also has limitations. 
    For example, when attempting to identify a friend in a public setting, we may occasionally mistake other individuals for them. 
    A device could adopt a comparable approach, informing the user when it is uncertain. 
    For instance, it could provide an indication of a potential friend by tagging a person with a label such as "Maybe James Parker."
    This approach will enable the device to be transparent in uncertainty.
    Thus, researchers in AT could collaborate with people with CVI to develop methods that take account of model uncertainty without inducing sensory overload.
\end{itemize}

\subsubsection*{\textbf{Visual augmentations:}}
Visual augmentations refer to the methods by which VBAT devices enhance users' vision and is fundamental for all the challenges discussed in the study.
Previous research has explored these solutions in various domains such as AR/VR research \cite{orlosky2014fisheye, boyd2022manipulating, boyd2023global, hong2024visual, sutton2022look, veas2011directing, mendez2010focus}, game design \cite{dillman2018visual}, computer vision \cite{zhang2021star, kuo1995real, haritaoglu2003real}, OLV \cite{pur2023use, zhao2015foresee, zhao2016cuesee} and 360-degree video \cite{grogorick2018comparison}.
The potential solutions discussed by the participants can be categorised into four primary visual augmentations: 

\begin{itemize}
    \item Object awareness and guidance. 
    Studies in AR/VR have compared the effectiveness of visual guidance techniques \cite{sutton2022look} and also advocated for using flicker on objects for guidance in AR \cite{sutton2024flicker}.  
    However, P6 raised concerns about the potential risk of flashing lights triggering epilepsy.     
    \item Controlling brightness. Studies from computer vision have delved into algorithms for image enhancement in low-light conditions \cite{ying2017new, guo2023low, yang2019biological}. 
    Additionally, in AR/VR research, the concept of visual noise cancellation has been explored to mitigate the effects of bright lights and glare \cite{hong2024visual}. 
    \item Modifying the visual field. Numerous studies have investigated methods in the field of OLV \cite{orlosky2014fisheye, peli2001vision, peli2007applications, vargas2001p}. 
    These methods include using fish-eye lenses within AR headsets \cite{orlosky2014fisheye} and employing vision multiplexing, which involves minifying the contours of a wide field and presenting them over the user's functional field of view \cite{peli2001vision, peli2007applications, vargas2001p}. 
    Additionally, there are approaches to modulate peripheral vision by reduce motion sickness \cite{zhang2022programmable} and peripheral movement \cite{koshi2019augmented}.
    \item Enhancing text. Techniques from computer vision \cite{zhang2021star, kuo1995real, haritaoglu2003real} and OLV \cite{wolffsohn2007image, zhao2015foresee} has explored multiple ways including real-time image enhancement techniques for magnification, contrast enhancement, edge enhancement, and black and white reversal.
\end{itemize}

Langlotz et al. \cite{langlotz2024design} presented a design space for vision augmentations. 
Their framework and findings can inform future development and enable consistent categorisation of existing approaches especially VBAT devices.
However, research is needed to assess the applicability of these visual augmentations for people with CVI.

\subsubsection*{Practicality:} 
Visual augmentations demonstrate better practicality compared to contextual understanding.
However, two limitations stem from current HMDs:

\begin{itemize}
    \item Access to video feed on HMDs.
    While optical see-through HMDs, such as the Microsoft Hololens \cite{MicrosoftHoloLens}, can seamlessly integrate virtual objects into the real world, most image enhancement or visual field augmentation techniques rely on the device's ability to manipulate the user's perception through its camera feed.
    Unfortunately, many commercially available devices restrict access to their cameras, such as the Meta Quest Pro \cite{meta_quest_pro}, and Meta Quest 3 \cite{meta_quest_3}. 
    Past research has employed various methods to circumvent these restrictions, such as mounting external cameras on the headsets \cite{visual_noise} or recording through HMDs with cameras \cite{sutton2024flicker}. 
    Nonetheless, these workarounds limit the ability to test these devices in real-world settings, constraining researchers' efforts. 
    The latest software for the Apple Vision Pro \cite{apple_vision_pro} has opened up camera access with its enterprise API, indicating a promising direction for the future of these devices.
    \item Device ergonomics. As described by participants, one of the main considerations is the device ergonomics such as lightweight design and being discreet.
    Many of the devices on the market are bulky and impractical for day to day use, especially outdoors.
\end{itemize}

However, recent advances in AR and VR have introduced devices like the Varjo XR-4 \cite{Varjo_2024}, which aim to address these limitation. 
As a result, it is expected that more compact and affordable options will become available in the market, making these solutions more feasible. 
Therefore, while many of the discussed visual augmentations show promise, they still face limitations due to HMDs.

\subsubsection*{\textbf{Auditory augmentations:}}
Auditory augmentations refer to the methods by which VBAT devices use audio to enhance the users visual experience. 
The potential solutions discussed by participants can be categorised into two main auditory augmentations.
\begin{itemize}
    \item Spatial audio for directional sound. 
    This solution was mainly discussed as way to address C1 and C2.
    Several studies in AT for OVI have explored the use of spatial audio to aid in object localisation \cite{lock2020experimental, raina2023pointing}, navigation \cite{massiceti2018stereosonic}, and hazard detection \cite{fauzul2021navigation, may2020spotlights}. 
    Careful incorporation of these techniques alongside visual augmentations can facilitate quicker object localisation for participants.

    \item Voice isolation with background noise cancellation. 
    This was discussed as a way to address C5.
    Research in signal processing has investigated algorithms for voice isolation using machine learning \cite{shihab2023voice, kumarreview} and background noise cancellation \cite{cowan2020real, porr2022real, sambur1978adaptive}. 
    Integrating these with visual augmentations such as highlighting can assist people with CVI in reducing mental overload when attempting to focus on an object or person.

\end{itemize}

\subsubsection*{Practicality:} 
The audio augmentations are the most practical among the three and are ready for use.
However, further research is needed to explore the impact of audio on sensory overload, as noted by participants in the study.

\subsection{What are the Considerations for Researchers?}

Our research has revealed an understudied area at the intersection of VBAT and the unique requirements of people with CVI.
This gap can be bridged in two main ways:
\subsubsection*{\textbf{Inclusion of CVI participants in OLV focused assistive device research:}}
Given the shared reliance on the visual modality and the availability of numerous assistive devices for OLV \cite{zhao2015foresee, lang2020augmented, zhao2016cuesee, zhao2019designing, zhao2020effectiveness}, evaluating their effectiveness for people with CVI is imperative.
We recommend separate reporting of findings and observations for people with CVI. 
Moreover, given the potential permanent alterations in visual pathways due to neuroplasticity, the long-term effects of using such assistive devices by people with CVI should be considered.

\subsubsection*{\textbf{Co-designing devices with a focus on CVI\@:}}
With the unique requirements of people with CVI such as vision-centric interaction and complexity impact on vision, it is essential to involve them from the requirements gathering/design to the evaluation stages.
Existing research in the field of HCI has consistently demonstrated the advantages of involving end-users in the design process, ranging from superior idea generation to shorter development cycles and increased user satisfaction \cite{steen2011benefits, kujala2003user, empathy}. 
Duckett \& Pratt \cite{participant_importance} also emphasised that the visually impaired community believe that researchers could make a substantial contribution by involving them in research design and practice.
Therefore, VBAT has a significant opportunity to involve people with CVI from the start when designing devices for them.

\section{Limitations and Future Work}
The analysis was constrained by the small dataset of 17 papers, limiting the scope of definitive conclusions. With only three papers specifically addressing VBAT, this significant lack of literature underscores the glaring research gap in this important research area.

The initial inclusion process in the scoping review was conducted primarily by a single researcher over two iterations. 
As a result, there is a possibility that some relevant works may have been overlooked.

Recruitment posed a notable challenge, particularly with adult participants with CVI. 
We observed two primary barriers: first, a lack of formal diagnosis, as detailed in the paper, and second, diagnoses of specific CVI conditions like Hemianopsia and Akinetopsia. 
Consequently, some people with CVI may have been excluded from participation due to diagnoses specific to their condition rather than an overarching formal CVI diagnosis.
Therefore, our study's participant pool was limited to just seven individuals, with only two diagnosed with cerebral palsy. 
This sample size may not suffice to yield conclusive findings, warranting future investigations with larger cohorts. 

This limitation of specific conditions extended to our scoping review as well, particularly in our keyword selection. 
While some papers discussed conditions related to CVI, such as Hemianopsia \cite{amini2020design, pelak2007homonymous, costela2018people} and Hemineglect \cite{karner2019effects, prangrat2000impact}, they did not explicitly mention CVI. 
Hence, it's plausible that relevant papers may have been overlooked due to the absence of the terms related to CVI.

In future studies, we plan to adopt a human-centred approach, by employing co-design methodologies with people with CVI to develop solutions identified in this study.

\section{Conclusion}

In this paper, we conducted two studies to explore the current landscape of research at the intersection of CVI and VBAT, while also exploring the opportunities for VBAT devices with people with CVI. 
Study 1 entailed a scoping review of 17 papers, while Study 2 involved focus group discussions with 7 participants living with CVI.

Our findings revealed a significant research gap at the intersection of CVI and VBAT, with existing studies predominantly focused on understanding CVI rather than addressing assistive needs.
Subsequently, we identified 7 overarching challenges faced by people with CVI, spanning from object and people awareness to managing sensory overload during face-to-face conversations. 
Moreover, drawing insights from previous research and our focus group discussion, we delineated similarities and differences in the AT needs of people with CVI compared to those with OLV. 
Finally, we aimed to raises awareness of CVI in HCI and Assistive Technology (AT) Communities.

With recent technological advancements for VBAT and a wide-open field for research, now is the ideal time for researchers to create a substantial impact on people's lives.
As CVI emerges as a prominent vision impairment, we call upon researchers in the fields of HCI and AT to recognise and address this gap.

\vspace{3mm}

\section*{Positionality Statement}
The research team includes one researcher who lives with CVI and is an expert in education and rehabilitation for children with CVI. All but one researcher on the team have experience working closely with people who are blind or have low vision.
The two researchers who coded the transcripts include a PhD candidate with experience in machine learning and a late-career academic with expertise in human-centred research. 
Neither had lived experience of OVI or CVI.

%Our collaborative effort ensured that the research was conducted sensitively and appropriately within the context of the study.

\begin{acks}

We are grateful to the Monash Data Futures Institute for providing the funding that made this project possible.
We would like to extend our appreciation to our participants Christina Roma, Colin Cook, Dijana Kovacic, Helen MacNee, Marjorie Mc Wee and William Cozis for their invaluable contribution to our study.
Their active participation significantly enriched the study, and we are truly appreciative of their involvement.

\end{acks}

\bibliographystyle{ACM-Reference-Format}
\bibliography{sample-base}

\appendix
%TC:ignore
\newpage
\section{REVIEW PAPERS}
\begin{table}[!h]
    \caption{List of 17 papers from the scoping review with the extracted data: Paper, Year, Age Group (A - Adults only, C - Children only, B - Both, ? - Unclear, N - None), Age Range ([Youngest] - [Eldest], ? - Unclear, N - None), Type of Study (D - Diagnosis, U - Understanding of CVI, S - Simulation, A - Assistance, R - Rehabilitation), Involvement (D - Design Stage, E - Evaluation Stage, B - Both, N - None), Type of Technology} 
    \begin{tabular}{|l|l|l|l|l|l|p{2.8cm}|}
        \hline
        \textbf{Study} & \textbf{Year} & \rotatebox[origin=c]{90}{\space \textbf{Age Group}} & \rotatebox[origin=c]{90}{\space \textbf{Age Range}} & \rotatebox[origin=c]{90}{\space \textbf{Type of Study}}  & 
        \rotatebox[origin=c]{90}{\space \textbf{Involvement} \space} & \textbf{Type of Technology} \\ \hline
        \cite{birnbaum2015enhancing} & 2015 & N & N & A & N & Computer Vision,  Image Enhancement,  Augmented Reality, Mixed Reality \\ \hline
        \cite{bennett2018virtual} & 2018 & N & 14-28 & U & E & Virtual Reality \\ \hline
        \cite{al20183d} & 2018 & N & N & S & N & Computer Vision, Artificial Intelligence,   Machine Learning, Virtual Reality \\ \hline
        \cite{bennett2018assessing} & 2018 & B & 14-28 & U & E & Virtual Reality \\ \hline
        \cite{bennett2020alpha} & 2020 & B & 14-25 & U & E & Virtual Reality \\ \hline
        \cite{lorenzini2021personalized} & 2021 & ? & ? & A, R & E & Computer Vision,  Image Enhancement,   Machine Learning \\ \hline
        \cite{pamir2021visual} & 2021  & C & 17 & U & E & Virtual Reality \\ \hline
        \cite{pitt2023strategies} & 2021 & N & N & A & N & Computer Vision,  Image Enhancement,  Machine Learning \\ \hline
        \cite{bennett2021visual} & 2021 & B & 15-22 & U & E & Virtual Reality \\ \hline
        \cite{federici2022altered} & 2022 &  B & 14-21 & U & E & Virtual Reality \\ \hline
        \cite{bambery2022virtual} & 2022 &  ? & ? & U, D & E & Virtual Reality \\ \hline
        \cite{zhang2022assessing} & 2022 &  B & 7-20 & U & E & Virtual Reality \\ \hline
        \cite{manley2022assessing} & 2022 &  B & 11-20 & D & E & Virtual Reality \\ \hline
        \cite{soni2023convolutional} & 2023 & N & N & D & N & Computer Vision,   Artificial Intelligence,   Image Enhancement,   Machine Learning \\ \hline
        \cite{avramidis2024evaluating} & 2024 & C & 1-12 & U & E & Computer Vision, Machine Learning \\ \hline
        \cite{da2024abstract} & 2024 & ? & ? & U & E & Virtual Reality \\ \hline
        \cite{tambala2024abstract} & 2024 & ? & ? & U & E & Virtual Reality \\ \hline
        \end{tabular}
        \label{tab:full-paper-list} \\
\end{table}

% \section{SEARCH PHRASES}
% \begin{verbatim*}
% ("cerebral visual impairment" OR 
% "cortical visual impairment" OR "neurological vision impairment") 
% AND
% ("computer vision" OR "artificial intelligence" OR
% "machine learning" OR "image enhancement" OR
% "augmented reality" OR "virtual reality"
% OR "mixed reality")
% \end{verbatim*}  
%TC:endignore

\end{document}